\documentclass[fleqn,usenatbib]{mnras}

\usepackage{newtxtext,newtxmath}

\usepackage[T1]{fontenc}
\usepackage{ae,aecompl}

\usepackage{graphicx}	
\usepackage{amsmath}	
\usepackage{amssymb}	
\usepackage{multicol}        
\usepackage{bm}		
\usepackage[a4paper]{geometry}
\usepackage{booktabs}
\usepackage{float}
\usepackage{placeins}
\usepackage{afterpage}

\title[INPOP17a and fundamental physics tests]{The new lunar ephemeris INPOP17a and its application to fundamental physics}

\author[V. Viswanathan]{
V. Viswanathan,$^{1,2}$\thanks{E-mail: viswanathan@geoazur.unice.fr}
A. Fienga,$^{1,2}$
O. Minazzoli,$^{3,4}$
L. Bernus,$^{2}$
J. Laskar,$^{2}$
M. Gastineau$^{2}$
\\
$^{1}$AstroG\'eo, G\'eoazur - CNRS UMR 7329 - Observatoire de la C\^ote d'Azur, 250 Rue Albert Einstein, 06560 Valbonne\\
$^{2}$ASD, IMCCE - CNRS UMR 8028 - Observatoire de Paris, 61 Avenue de l'Observatoire, 75014 Paris\\
$^{3}$Centre Scientifique de Monaco, 8 Quai Antoine Ier, MC 98000, Monaco\\
$^{4}$Artemis - CNRS UMR 7250 - Observatoire de la C\^ote d'Azur, 96 Boulevard de l'Observatoire, 06300 Nice
}
\date{Accepted XXX. Received YYY; in original form ZZZ}

\pubyear{2017}

\begin{document}
\label{firstpage}
\pagerange{\pageref{firstpage}--\pageref{lastpage}}
\maketitle

\begin{abstract}
We present here the new INPOP lunar ephemeris, INPOP17a. This ephemeris is obtained through the numerical integration of the equations of motion and of rotation of the Moon, fitted over 48 years of Lunar Laser Ranging (LLR) data. We also include the 2 years of infrared (IR) LLR data acquired at the Grasse station between 2015 and 2017. Tests of the universality of free fall are performed. We find no violation of the principle of equivalence at the $10^{-14}$ level. A new interpretation in the frame of dilaton theories is also proposed.
\end{abstract}


\section{Introduction}
The Earth-Moon system is an ideal tool for carrying out tests of general relativity and more particularly the test of the universality of free fall \cite[]{1968PhRv..169.1017N,1996ApJ...459..365A}. Since 1969, the lunar laser ranging (LLR) observations are obtained on a regular basis by a network of laser ranging stations \cite[]{Faller1969,Bender1973}, and currently with a millimeter-level accuracy \cite[]{Samain1998,Murphy2013a}. Thanks to this level of accuracy at the solar system scale, the principle of the universality of free fall (UFF) can in theory be tested. However, at these accuracies (of 1 cm or below), the tidal interactions between the Earth and the Moon are complex to model, especially when considering that the inner structure of the Moon is poorly known \cite[]{Wieczorek2007,Williams2015}. This explains why the UFF test is only possible after an improvement of the dynamical modeling of the Earth-Moon interactions.

Recently, thanks to the GRAIL mission, an unprecedented description of the lunar shape and its variations were obtained for the 6 months of the duration of the mission \cite[]{2014GeoRL..41.1452K}. This information is crucial for a better understanding of the dissipation mechanism over longer time span \cite[]{2016GeoRL..43.8365M,2015GeoRL..42.7351M,Williams2015}. Furthermore, since 2015, the Grasse station which produces more than 50 \% of the LLR data, has installed a new detection path at 1064 nm (IR) ranging wavelength leading to a significant increase of the number of observations and of the signal to noise ratio \cite[]{Courde2017}.

Together with these new instrumental and GRAIL developments, the Moon modeling of the INPOP planetary ephemeris was improved. Since 2006, INPOP has become a reference in the field of the dynamics of the solar system objects and in fundamental physics \cite[]{fienga:2011cm,Fienga2016a}.

The INPOP17a version presented here also benefits some of the planetary improvements brought by the use of updated Cassini deduced positions of Saturn. The planetary and lunar Chebyshev polynomials built from INPOP17a have been made available on the INPOP website together with a detailed technical documentation \cite[]{Viswanathan2017a}.

Since 2010, thanks to the millimeter-level accuracy of the LLR measurements and the developments in the dynamical modeling of the Earth-Moon tidal interactions, differences in acceleration of Earth and Moon in free fall towards the direction of the Sun could reach an accuracy of the order of 10$^{-14}$ \cite[]{2010LRR....13....7M,Williams2012}. With the improvement brought by GRAIL, addition of IR LLR observations and the recent improvement of the dynamical modeling of INPOP17a, one can expect to confirm or improve this limit.

In this paper, we first present (see section \ref{npt}) the statistics related to the IR dataset obtained at the Grasse station since 2015. In section (\ref{dynmodel}), we introduce the updated dynamical model of the Moon as implemented in the INPOP planetary ephemeris including contributions from the shape of the fluid core. In section (\ref{appendixB}), we explain how we use the IR data to fit the lunar dynamical model parameters with the GRAIL gravity field coefficients as a supplementary constraint for the fluid core description.

Finally in section (\ref{fundphy}) we describe how we test the UFF and give new constraints. In addition, we present a generalization of the interpretation in terms of gravitational to inertial mass ratios of UFF constraints, based on recent developments in dilaton theories \cite[]{hees:2015ax,minazzoli:2016pr}. Hinged on this generalization, we deduce that from a pure phenomenological point of view, one cannot interpret UFF violation tests in the Earth-Moon system as tests of the difference between gravitational and inertial masses only.

\section{Lunar ephemerides}
The new INPOP planetary ephemerides INPOP17a \citep[]{Viswanathan2017a} is fitted to LLR observations from 1969 to 2017, including the new IR LLR data obtained at the Grasse station.

\begin{figure*}
    \centering
    \includegraphics[width=1.0\textwidth]{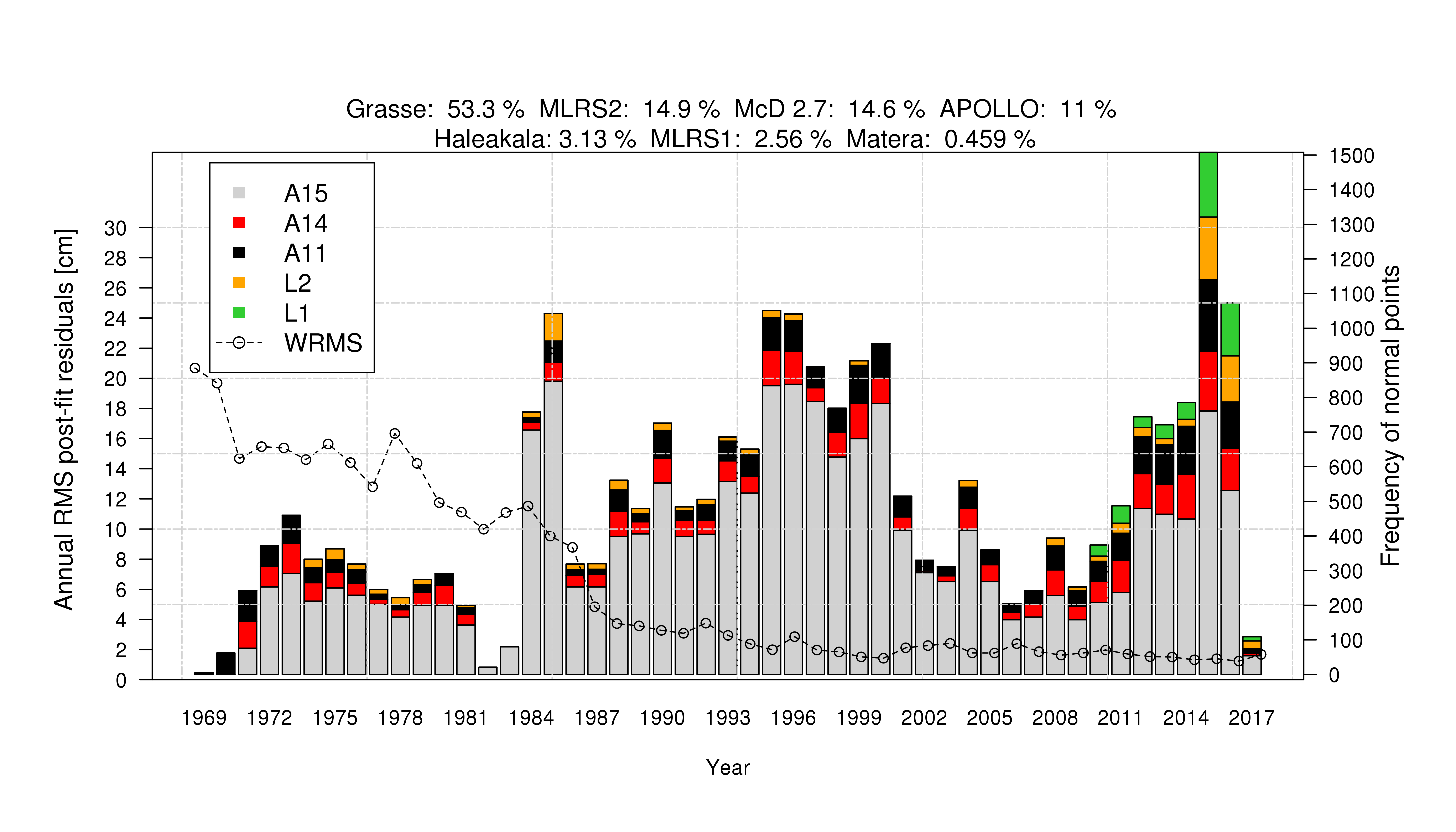}
    \caption{Histogram of annual frequency of LLR data with relative contribution from each LRR array including Grasse IR (1064 nm) observations. Points indicate the annual mean of post-fit residuals (in cm) obtained with INPOP17a. The dominance of range observations to A15 is evident. A change can be noticed after 2014 due to the contribution from IR at Grasse. }
    \label{NPT_dist}
    \end{figure*}

\subsection{Lunar Laser Ranging}
  \label{npt}
  The principle of the LLR observations is well documented \citep[]{Murphy2013a,Murphy2012}. Besides the lunar applications, the laser ranging technique is still intensively used for tracking Earth orbiting satellites, especially for very accurate orbital \citep[]{2013MNRAS.432.2591P,2015CQGra..32o5012L} and geophysical studies \citep[]{2013GeoRL..40.4662M,Jeon2011}.
  
 \begin{figure*}
    \centering
         \includegraphics[width=0.35\textwidth]{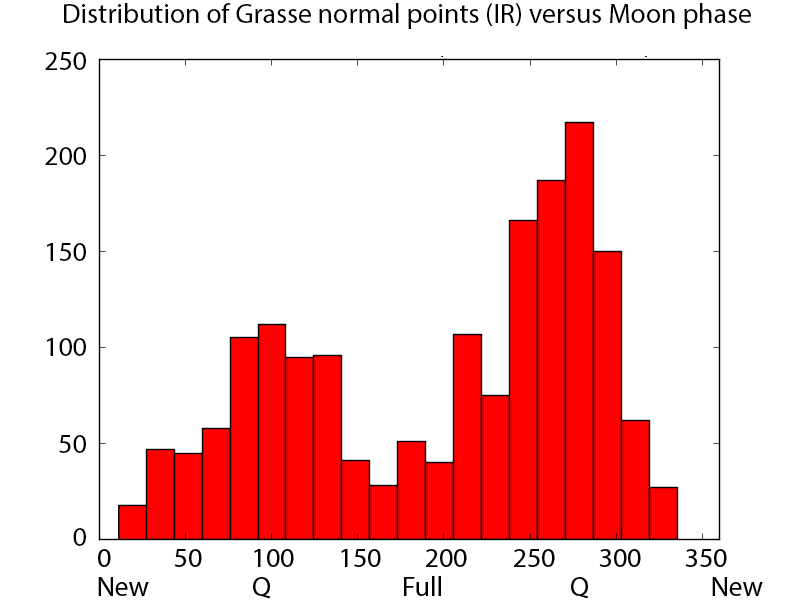}\includegraphics[width=0.35\textwidth]{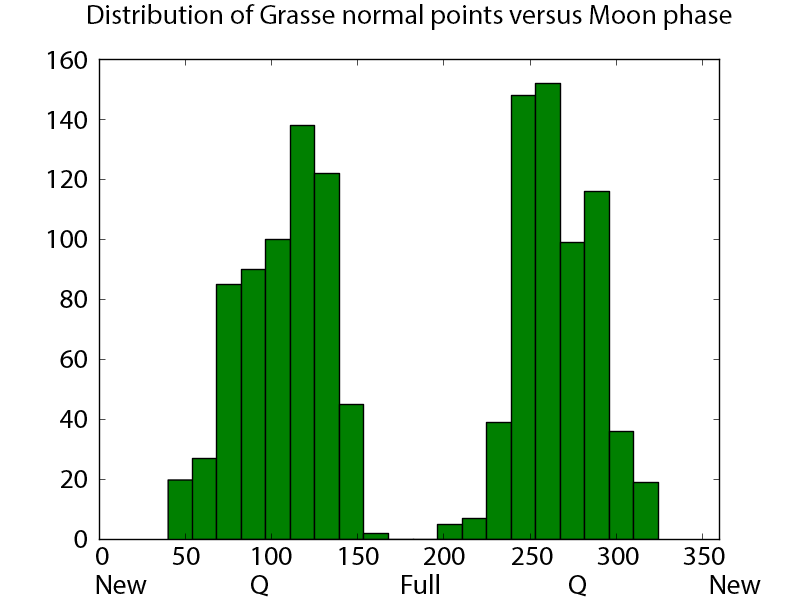}\includegraphics[width=.35\textwidth]{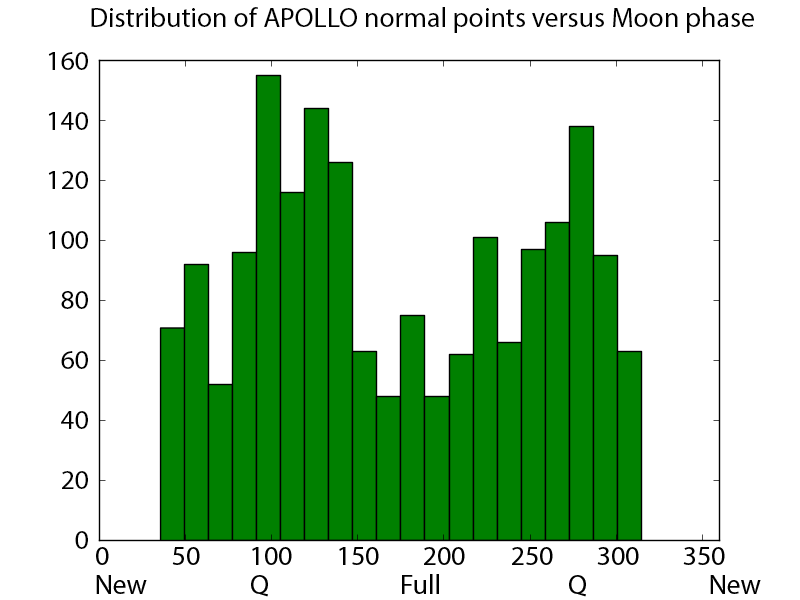}
          \caption{Histogram of synaptic distribution of normal points obtained at Apache Point (right-hand side), at the Grasse station from 2012 from 2014 at 542 nm (center) and from 2014 to 2016 at 1064 nm (left-hand side). Q indicates the quarter Moon phase.}
           \label{NPT_syn_hist}
    \end{figure*}

\begin{figure}
      \centering
      \includegraphics[width=0.5\textwidth]{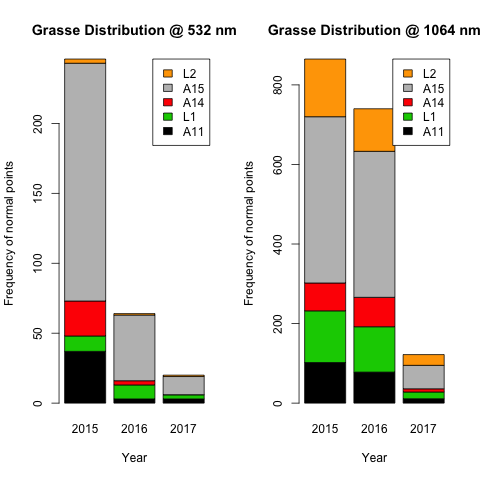}
      \caption{Grasse reflector wise distribution at 532 nm and 1064 nm from 2015 to 2017.}
      \label{NPT_REFL}
\end{figure}

Non-uniform distributions in the dataset are one contributor to correlations between solution parameters \citep[]{Williams2009}. Like one can see on Fig. (\ref{NPT_dist}), Fig. (\ref{NPT_syn_hist}) and Fig. (\ref{NPT_REFL}), about 70 \% of the data are obtained after reflection on A15 reflector and on an average 40 $\%$ of the data are acquired at 30$^{\circ}$ apart from the quarter Moons. 
    
    In this study, we show how the IR LLR observations acquired at the Grasse station between 2015 and 2017 (corresponding to 7 \% of the total LLR observations obtained between 1969 and 2017 from all known ILRS ground stations) can help to reduce the presence of such heterogeneity.
      
      \subsubsection{Spatial distribution}
      \label{sel_bias}
       Statistics drawn from the historical LLR dataset (1969-2015) show an observer bias to range to the larger Apollo reflector arrays (mainly A15). This trend (see Fig. \ref{NPT_dist} and Fig. \ref{NPT_REFL}) is also present on statistics taken during time periods after the re-discovery of Lunokhod 1 by \cite{Murphy2011}.
       This is due to the higher return rate and thermal stability over a lunar day on the Apollo reflectors, thereby contributing to the higher likelihood of success. 
       
       With the installation of the 1064 nm detection path (see Fig. \ref{NPT_REFL}), as explained in \cite{Courde2017}, the detection of photon reflected on all reflectors is facilitated, especially for Lunokhod 2 (L2): about 17 \% of IR data are obtained with L2 when only 2 \% were detected at 532 nm.
 
        Owing to the spatial distribution of the reflectors on the Moon, Apollo reflectors offer principally longitude libration sensitivity at the Moon equator, whereas Lunokhod reflectors offer sensitivity both in the latitude and longitude libration of the Moon.  The heterogeneity in the reflector-wise distribution of LLR data affects then the sensitivity of the lunar modeling adjustment \cite[]{2016EGUGA..1813995V}.
         By acquiring a better reflector-wise sample, IR contributes to improve the adjustment of the Moon dynamical and rotational modeling (see section \ref{appendix_est}). 

\subsubsection{Temporal distribution}
      \label{phas_bias}
    The full and new Moon periods are the most favorable for testing gravity, as the gravitational and tidal effects are maximum. This was partially demonstrated by \cite{1998CQGra..15.3363N}.
      On Fig. (\ref{NPT_syn_hist}) are plotted the distributions of normal points relative to the synaptic angle for APOLLO and Grasse station obtained at 532 nm and 1064 nm. When for the APOLLO data sets the distribution of normal points around quarter Moons (15$^{\circ}$ before and after 90$^{\circ}$ and 270$^{\circ}$) correspond to about 25 \% of the full data sample, almost 45 \% of the Grasse 532 nm data sample is obtained away from the full and new Moon periods. This can be explained by two factors:
      
      \begin{enumerate}[a.]
      \item New Moon phase\\As the pointing of the telescope onto the reflectors is calibrated with respect to a nearby topographical feature on the surface of the Moon, the pointing itself becomes a challenge when the reference points lie in the unlit areas of the Moon. Also, as the New Moon phase occurs in the daylight sky, the noise floor increases and the detector electronics become vulnerable due to ranging at a very close angle to the Sun \cite[]{Courde2017,Williams2009}.

      \item Full Moon phase\\During this phase, thermal distortions remain as the primary challenge, arising due to the over-head Sun heating of the retro-reflector arrays. This induces refractive index gradients within each corner cube causing a spread in the return beam, which makes detection more difficult. The proportion of this effect is partially linked to the thermal stability of the arrays. Since the A11, A14 and A15 arrays have a better thermal stability compared to the L1 and L2 arrays \cite[]{Murphy2014}, observations to the latter become sparse during the full Moon phase.
    \end{enumerate}

 Despite these challenges, LLR observations during the above mentioned phases of the Moon have been acquired with the IR detection. 
      For the first two years of 1064 nm detection path at the Grasse station, about 32 \% of observations were indeed obtained at 30$^{\circ}$ apart from the Moon quarters, increasing by 10 \% the portion of data sample close from the most favorable periods for tides and UFF studies.
      
      This is primarily achieved due to the improved signal to noise ratio resulting from an improved transmission efficiency of the atmosphere at the IR wavelength of 1064 nm. In addition, high precision data have also been acquired on the two Lunokhod reflector arrays during full Moon phase.

      In section (\ref{fundphy}), we will see how the IR LLR data help to improve the results related to the UFF tests.

      \subsubsection{Observational Accuracy of the LLR observations}
      \label{accuracy}
\begin{figure}
      \centering
      \includegraphics[width=0.4\textwidth]{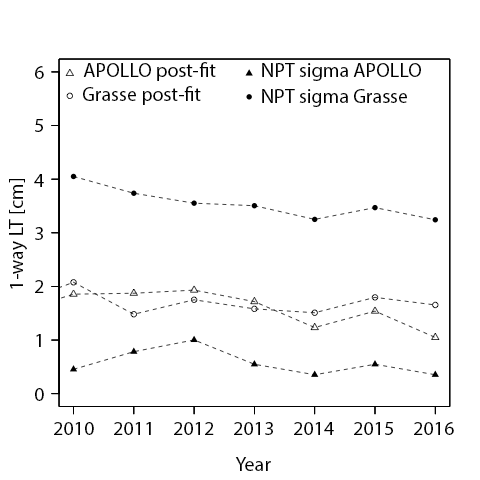}
      \caption{APOLLO and Grasse LLR observations in terms of i) observational accuracy as given by the annual mean of normal point uncertainty (converted from ps to 1-way light time (LT) in cm) and ii) annual weighted root mean square of post-fit residuals (1-way LT in cm) obtained with INPOP17a.}
      \label{NPT_accuracy}
      \end{figure}

      APOLLO observations are obtained with a 3.5 m telescope (under time sharing) at the Apache Point Observatory, while Grasse observations are obtained with a 1.5 m telescope dedicated for SLR and LLR. A larger aperture is beneficial for statistically reducing the uncertainty of the observation \cite[]{Murphy2013a}, which translates to millimeter level accuracies for APOLLO. One can notice in Fig. (\ref{NPT_accuracy}) that the current lunar ephemerides have a post-fit residual scatter (RMS) of about 1-2 cm for the recent observations while the LLR normal point accuracy is given to be at least two times smaller. This calls for an improvement of the Earth-Moon dynamical models within highly accurate numerically integrated ephemerides (see section \ref{appendix_est}). 

\subsection{Lunar Dynamical Model}
  \label{dynmodel}
      \subsubsection{Lunar orbit interactions}
    \label{lunorbit}
    In our model, we include the following accelerations perturbing the Moon's orbit:
      \begin{enumerate}
      \item Point mass mutual relativistic interactions, in the parametrized post-Newtonian formalism, from the Sun, planets and asteroids through \citet[Eqn. 27]{Folkner2014};
      \item Extended bodies mutual interactions, through \citet[Eqn. 28]{Folkner2014}, which include :
      \begin{itemize}
        \item the interaction of the zonal harmonics of the Earth through degree 6;
        \item the interaction between zonal, sectoral, and tesseral harmonics of the Moon through degree 6 and the point mass Earth, Sun, Jupiter, Saturn, Venus and Mars;
        \item the interaction of degree 2 zonal harmonic of the Sun.
      \end{itemize}
      \item Interaction from the Earth tides, through \citet[Eqn. 32]{Folkner2014}
      \end{enumerate}
      The tidal acceleration from the tides due to the Moon and the Sun are separated into three frequency bands (zonal, diurnal and semi-diurnal). Each band is represented by a potential Love number $k_{2m,E}$ with a matching pair of time delays $\tau_{Xm,E}$ (where subscript $X$ is either associated with the daily Earth rotation $\tau_{Rm,E}$ or orbital motion $\tau_{Om,E}$) to account for frequency dependent phase shifts from an anelastic Earth with oceans. Here the time delay represents the phase lag induced by the tidal components. Although the time delay method inherently assumes that the imaginary component of $k_{2m,E}$ varies linearly with frequency, it reduces the complexity of the dynamical model. The diurnal $\tau_{R1,E}$ and semi-diurnal $\tau_{R2,E}$ are included as solution parameters in the LLR analysis, while model values for potential Love numbers for a solid Earth are fixed to that from \citet[Table 6.3]{IERS2010} followed by corrections from the ocean model FES2004 \cite[]{Lyard2006}. A detailed explanation about the most influential tides relevant to the Earth-Moon orbit integration can be found in \citet[Table 6]{Williams2016Secular}.

\subsubsection{Lunar orientation and inertia tensor}
\begin{enumerate}
\item Lunar frame and orientation\\The mantle coordinate system is defined by the principal axes of the undistorted mantle, whose moments of inertia matrix are diagonal. The time varying mantle Euler angles ($\phi$$_m$($t$),$\theta$$_m$($t$),$\psi$$_m$($t$)) define the orientation of the principal axis (PA) frame with respect to the inertial ICRF2 frame (see \cite{Folkner2014} for details). The time derivatives of the Euler angles are defined through \citet[Eqn. 14]{Folkner2014}.

      \item Lunar moment of inertia tensor \\
      The undistorted total moment of inertia of the Moon \~{I}$_T$ is given by:
      \begin{eqnarray}\label{MoI_Moon}
        \tilde{I}_T &=&  \dfrac{\tilde{C}_T}{m_{M}R^{2}_{M}}\begin{bmatrix}
                                                         1 &0 &0 \\
                                                         0 &1 &0 \\
                                                         0 &0 &1\end{bmatrix}\nonumber \\
                                                         &&+
          \begin{bmatrix}
                  \tilde{C}_{2,0,M} - 2\tilde{C}_{2,2,M}      &0                                          &0    \\
                  0                                           &\tilde{C}_{2,0,M} + 2\tilde{C}_{2,2,M}     &0    \\
                  0                                           &0                                          &0
          \end{bmatrix}
      \end{eqnarray}

      where \~{C}$_{n,m,M}$ is the unnormalized degree n, order m of the Stokes coefficient $C_{n,m}$ for the spherical harmonic model of the undistorted Moon and $\tilde{C}_T$ is the undistorted polar moment of inertia of the Moon normalized by it's mass $m_M$ and radius squared $R^{2}_M$. Through Eqn. (\ref{MoI_Moon}), we are able to directly use the undistorted value of $C_{22}$ \cite[]{Manche2011} from GRAIL derived spherical harmonic model of \cite{Konopliv2013}.

      The moment of inertia of the fluid core $I_c$ is given by:
      \begin{subequations}
      \begin{equation}\label{MoI_c}
        I_c =    \alpha{_c}\tilde{C_T}\begin{bmatrix}
                  1-f_c     &0         &0    \\
                  0         &1-f_c     &0    \\
                  0         &0         &1
                                  \end{bmatrix} = \begin{bmatrix}
                                        A_c       &0         &0    \\
                                        0         &B_c       &0    \\
                                        0         &0         &C_c
                                        \end{bmatrix}
      \end{equation}
      where $\alpha_c$ is the ratio of the fluid core polar moment of inertia $C_c$ to the undistorted polar moment of inertia of the Moon $C_T$, $f_c$ is the fluid core polar flattening and, $A_c$ and $B_c$ are the equatorial moments of the fluid core. This study assumes an axis-symmetric fluid core with A$_c$ = B$_c$.

      The moment of inertia of the mantle $I_m$ has a rigid-body contribution $\tilde{I}_m$ and two time varying contributions due to the tidal distortion of the Earth and spin distortion as given in \citet[Eqn. 41]{Folkner2014}. The single time delay model (characterized by $\tau_M$) allows for dissipation when flexing the Moon \cite[]{standish1992orbital,Williams2001,Folkner2014}. 

      \begin{equation}
        \tilde{I}_m = \tilde{I}_T - I_c
      \end{equation}

      \item Lunar angular momentum and torques\\
      The time derivative of the angular momentum vector is equal to the sum of torques ($N$) acting on the body. In the rotating mantle frame, the angular momentum differential equation for the mantle is given by:

      \begin{equation}\label{L_balance}
      \frac{d}{dt}I_m\omega_m + \omega_m \times I_m\omega_m = N
      \end{equation}
      where $N$ is the sum of torques on the lunar mantle from the point mass body $A$ ($N_{M,figM-pmA}$), figure-figure interaction between the Moon and the Earth ($N_{M,figM-figE}$) and the viscous interaction between the fluid core and the mantle ($N_{CMB}$).

      The motion of the uniform fluid core is controlled by the mantle interior, with the fluid core moment of inertia ($I_c$) constant in the frame of the mantle. The angular momentum differential equation of the fluid core in the mantle frame is then given by:
      \begin{equation}\label{L_balance_fcore}
      \frac{d}{dt}I_c\omega_c + \omega_m \times I_c\omega_c = -N_{CMB} 
      \end{equation}

      \begin{equation}\label{N_CMB}
      N_{CMB} = k_v \big( \omega_c - \omega_m \big) + \big( C_c - A_c \big) \big(\hat{z}_m \cdot \omega_c\big)\big(\hat{z}_m \times \omega_c \big)
      \end{equation}
      \end{subequations}
      where $k_v$ is the coefficient of viscous friction at the CMB and $\hat{z}_m$ is a unit vector aligned with the polar axis of the mantle frame. The second part on the right-hand side of Eqn. (\ref{N_CMB}) is the inertial torque on the axis-symmetric fluid core.
      \end{enumerate}

\subsection{Reduction model}
\label{appendixA}

    \begin{table*}
    \centering
    \caption{Comparison of post-fit residuals of LLR observations from ground stations with corresponding time span, number of normal points available, number of normal points used in each solution after a 3-$\sigma$ rejection filter. The WRMS (in~cm) is obtained with solutions INPOP13c (1969-2013) and INPOP17a (1969-2017). INPOP13c statistics are drawn from \protect\cite{inpop13c}.}
    \label{NPT_span}
    \begin{tabular}{clrrrrrr}
    \toprule
     &                                                                    &                &                   & \multicolumn{2}{c}{\textbf{INPOP13c}}  & \multicolumn{2}{c}{\textbf{INPOP17a}}     \\
     \textbf{Code}             &   \textbf{Station}    & \textbf{Time span}                                 & \textbf{Available}         & \textbf{Used}      & \textbf{WRMS}     & \textbf{Used}      & \textbf{WRMS}  \\ \cmidrule{5-8}
                  &                                            &                                  &                     &           &  [cm]    &           & [cm]         \\ \midrule
    70610         & APOLLO, NM, USA (group A)            & 2006 - 2010                  &  941              & 940       & 4.92     &  929          &  1.27        \\ 
    70610         & APOLLO, NM, USA (group B)            & 2010 - 2012                  &  506              & 414       & 6.61     &  486          &  1.95        \\ 
    70610         & APOLLO, NM, USA (re-group C)        & 2012 - 2013                  &  361              & 359       & 7.62     &  345          &  1.52        \\ 
    70610         & APOLLO, NM, USA (group D)            & 2013 - 2016                  &  832              &  -        &  -       &  800          &  1.15        \\ 
    01910         & Grasse, FR                                 & 1984 - 1986                              &  1187             & 1161      & 16.02    &  1161         &  14.01       \\ 
    01910         & Grasse, FR                                 & 1987 - 1995                              &  3443             & 3411      & 6.58     &  3407         &  4.11        \\ 
    01910         & Grasse, FR                                 & 1995 - 2006                              &  4881             & 4845      & 3.97     &  4754         &  2.86        \\ 
    01910         & Grasse, FR                                 & 2009 - 2013                              &  999              & 990       & 6.08     & 982           &  1.41        \\ 
    01910         & Grasse, FR                                 & 2013 - 2017                              &  3351             &  -        &  -       & 3320          &  1.51        \\ 
    56610         & Haleakala, HI, USA                         & 1984 - 1990                        &  770              & 739       & 8.63     & 728           &  4.80        \\ 
    07941         & Matera, IT                                 & 2003 - 2013                               &  83               & 70        & 7.62     & 37            &  2.37        \\ 
    07941         & Matera, IT                                 & 2013 - 2015                               &  30               &  -        & -        & 28            &  2.93        \\ 
    71110         & McDonald, TX, USA                          & 1969 - 1983                     &  3410             & 3302      & 31.86    & 3246          &  18.87       \\ 
    71110         & McDonald, TX, USA                          & 1983 - 1986                     &  194              & 182       & 20.60    & 148           &  16.77       \\ 
    71111         & MLRS1, TX, USA                             & 1983 - 1984                      &  44               & 44        & 29.43    & 44            &  32.73       \\
    71111         & MLRS1, TX, USA                             & 1984 - 1985                      &  368              & 358       & 77.25    & 356           &  62.58       \\
    71111         & MLRS1, TX, USA                             & 1985 - 1988                      &  219              & 207       & 7.79     & 202           &  11.07       \\ 
    71112         & MLRS2, TX, USA                             & 1988 - 1996                      &  1199             & 1166      & 5.36     & 1162          &  3.81        \\ 
    71112         & MLRS2, TX, USA                             & 1996 - 2012                      &  2454             & 1972      & 5.81     & 1939          &  3.72        \\ 
    71112         & MLRS2, TX, USA                             & 2012 - 2015                      &  17               &  -        & -        & 15            &  2.59        \\  \midrule
    \multicolumn{2}{r}{\textbf{TOTAL}}                   &  \textbf{1969 - 2017}                & \textbf{25289}                  & \textbf{20160} &     & \textbf{24089}         &              \\ \bottomrule
    \end{tabular}
    \end{table*}

  The reduction model for the LLR data analysis has been implemented within a precise orbit determination and geodetic software: GINS \cite[]{marty2011gins,viswanathan2015utilizing} maintained by space geodesy teams at GRGS/OCA/CNES and written in Fortran90. The subroutines for the LLR data reduction within GINS is vetted through a step-wise comparison study conducted among the LLR analysis teams in OCA-Nice (this study), IMCCE-Paris and IfE-Hannover, by using simulated LLR data and DE421 \cite[]{Folkner2009} as the planetary and lunar ephemeris. The modeling follows the recommendations of IERS 2010 \cite[]{IERS2010}. To avoid any systematics in the reduction model, the upper-limit on the discrepancy between the teams was fixed to 1 mm in one-way light time.

  From each normal point, the emission time (in UTC) and the round trip time (in seconds) are used to iteratively solve for the reflection time in the light-time equations. A detailed description is available in \citet[Section 8 \& 11]{Moyer2003} for a precise round-trip light-time computation. 
  
  A detailed description of the reduction model used for this study is provided in \cite{Manche2011}. 

 \subsection{Fitting procedure}
 \label{appendixB}

    For APOLLO station observations, scaling the uncertainties of the normal points depending on the change of equipments, or a change in the normal point computation algorithm, is advised (see \url{http://physics.ucsd.edu/~tmurphy/apollo/151201_notes.txt}). Unrealistic uncertainties present in observations from Grasse, McDonald MLRS2 and Matera between time periods 1998-1999, 1996 and 2010-2012 respectively, are rescaled. Additional details of the weighting scheme and the fitting procedure used for the construction of INPOP17a solution can be found in \cite{Viswanathan2017a}.
    A filtering scheme is enforced during the iterative fit of the parameters. At each iteration, the residuals are passed through a 3-$\sigma$ filter (where $\sigma$ is recomputed at each iteration).
      
    \subsubsection{Biases}
    \label{appendixB_bias}
    Changes in the ground station introduces biases in the residuals. These biases correspond either with a known technical development at the station (new equipment, change of optical fiber cables) or systematics. Any estimated bias can be correlated with a corresponding change in the ground station, provided the incidents have been logged. A list of known and detected biases are given in \cite{Viswanathan2017a}.

  \begin{figure}
    \centering
    \includegraphics[width=0.8\linewidth]{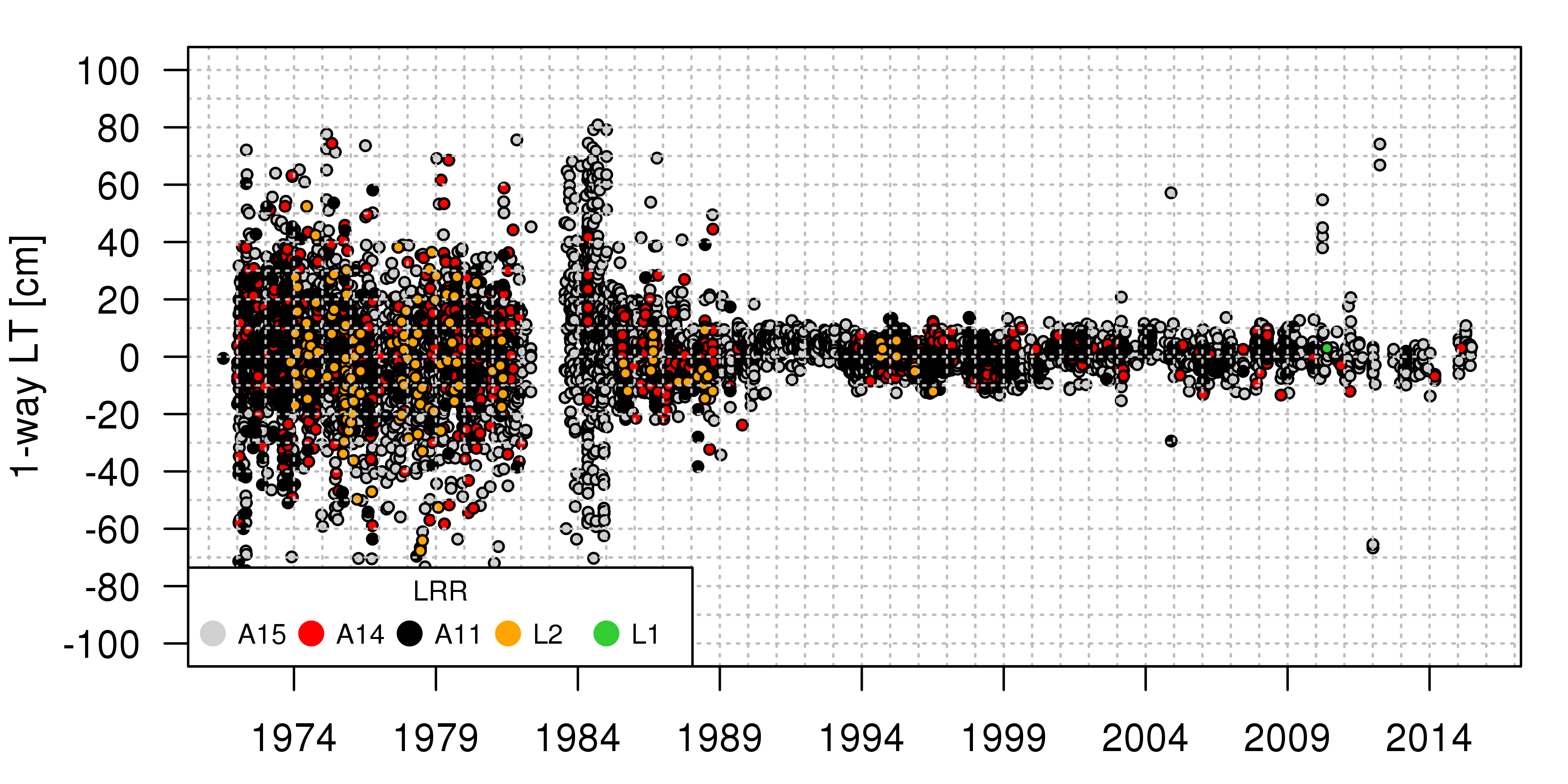}
    \caption{Post-fit residuals in (cm) vs time (year) obtained with INPOP$_{\textrm{G+IR}}$ specification (sec. \ref{appendix_est}) for McDonald, MLRS1, MLRS2, Haleakala and Matera stations}
    \label{fig:pf_rest} 
    \end{figure}

    \begin{figure}
    \centering
    \includegraphics[width=0.8\linewidth]{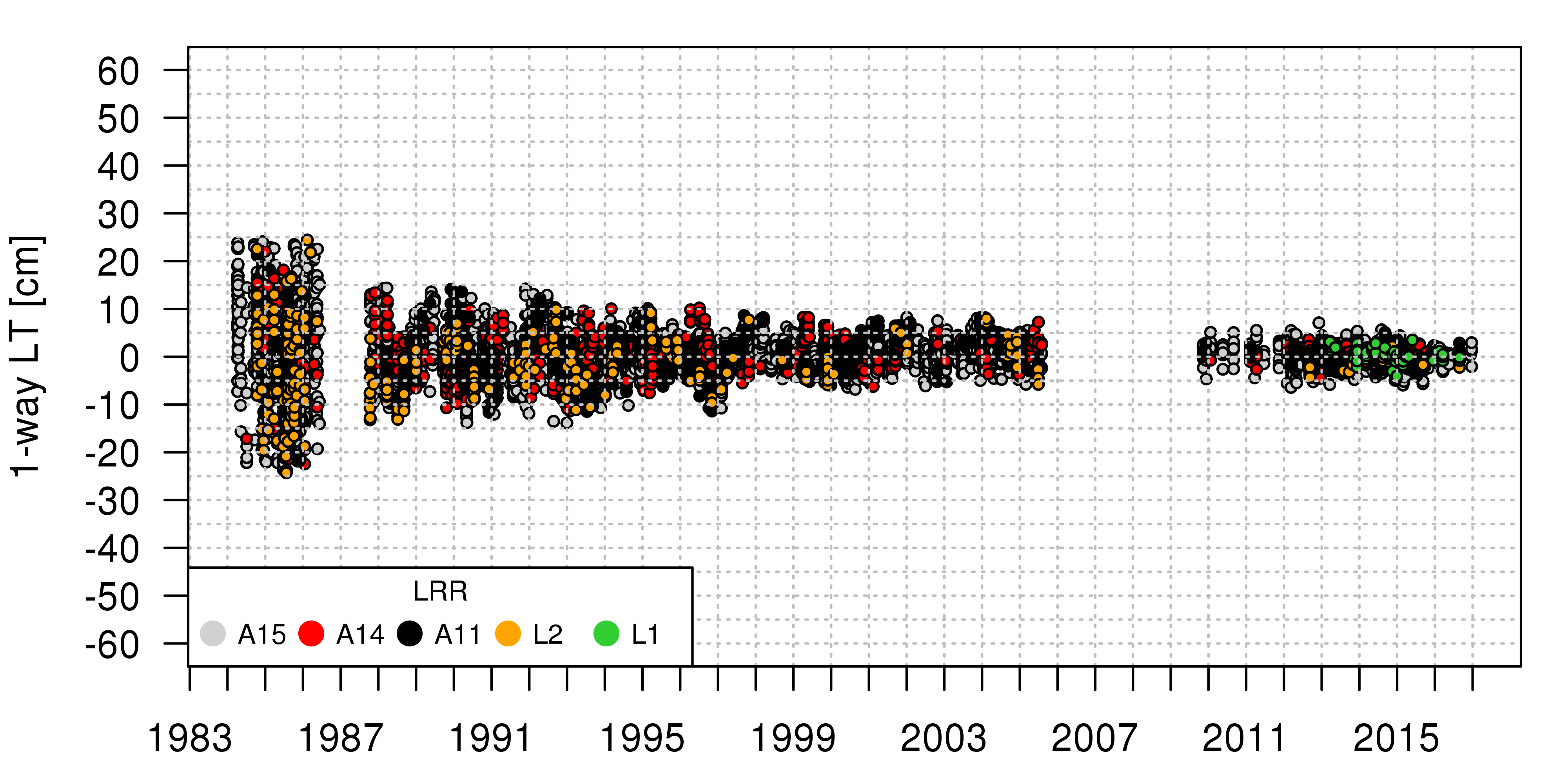}
    \caption{Post-fit residuals in (cm) vs time (year) obtained with INPOP$_{\textrm{G+IR}}$ specification (sec. \ref{appendix_est}) for GRASSE station with the Green wavelength}
    \label{fig:pf_Calern_gr}
   \end{figure}

    \begin{figure}
    \centering
    \includegraphics[width=0.8\linewidth]{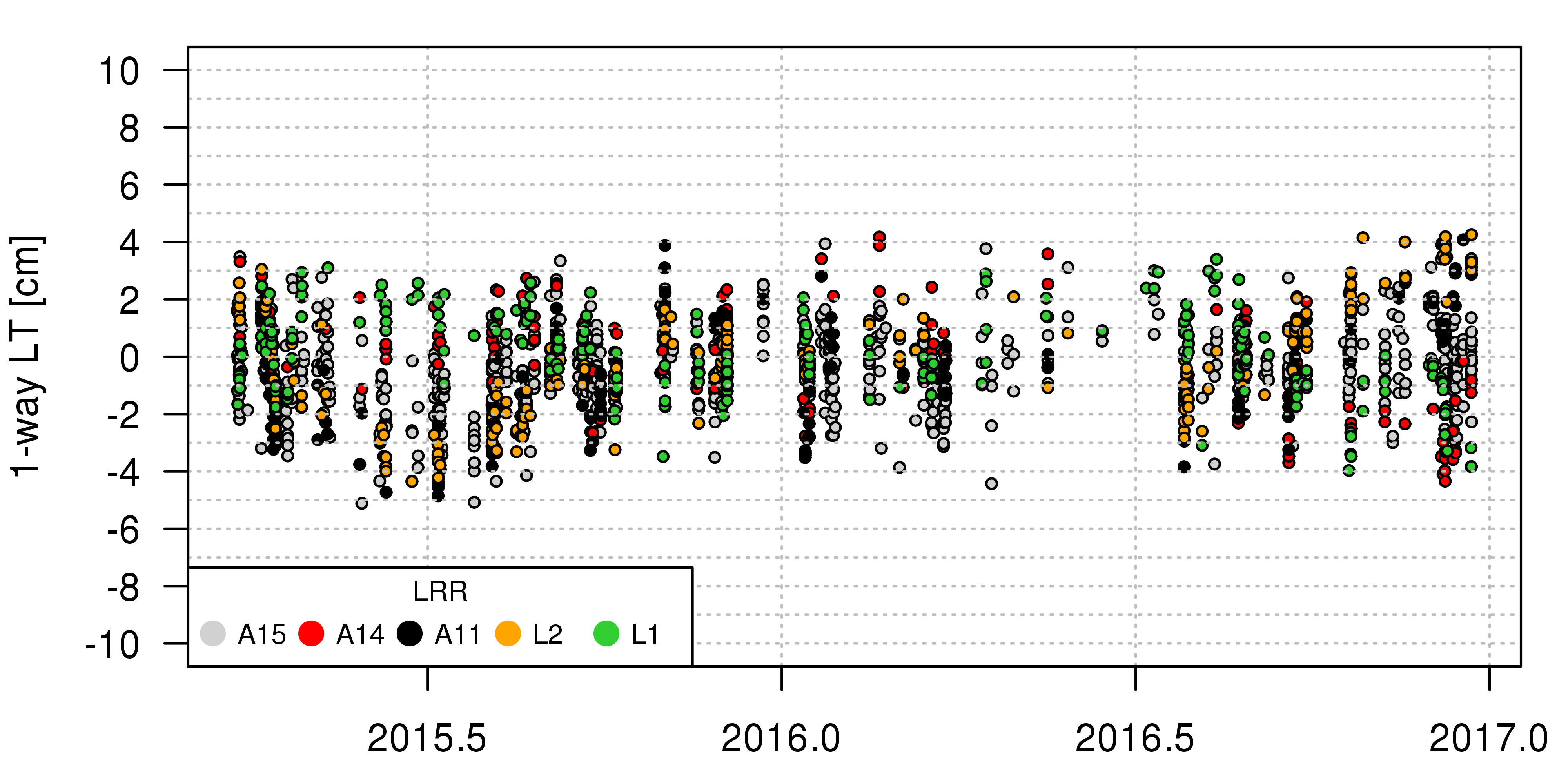}
    \caption{Post-fit residuals in (cm) vs time (year) obtained with INPOP$_{\textrm{G+IR}}$ specification (sec. \ref{appendix_est}) for GRASSE station with the IR wavelength}
    \label{fig:pf_Calern_ir}
    \end{figure}

    \begin{figure}
    \centering
    \includegraphics[width=0.8\linewidth]{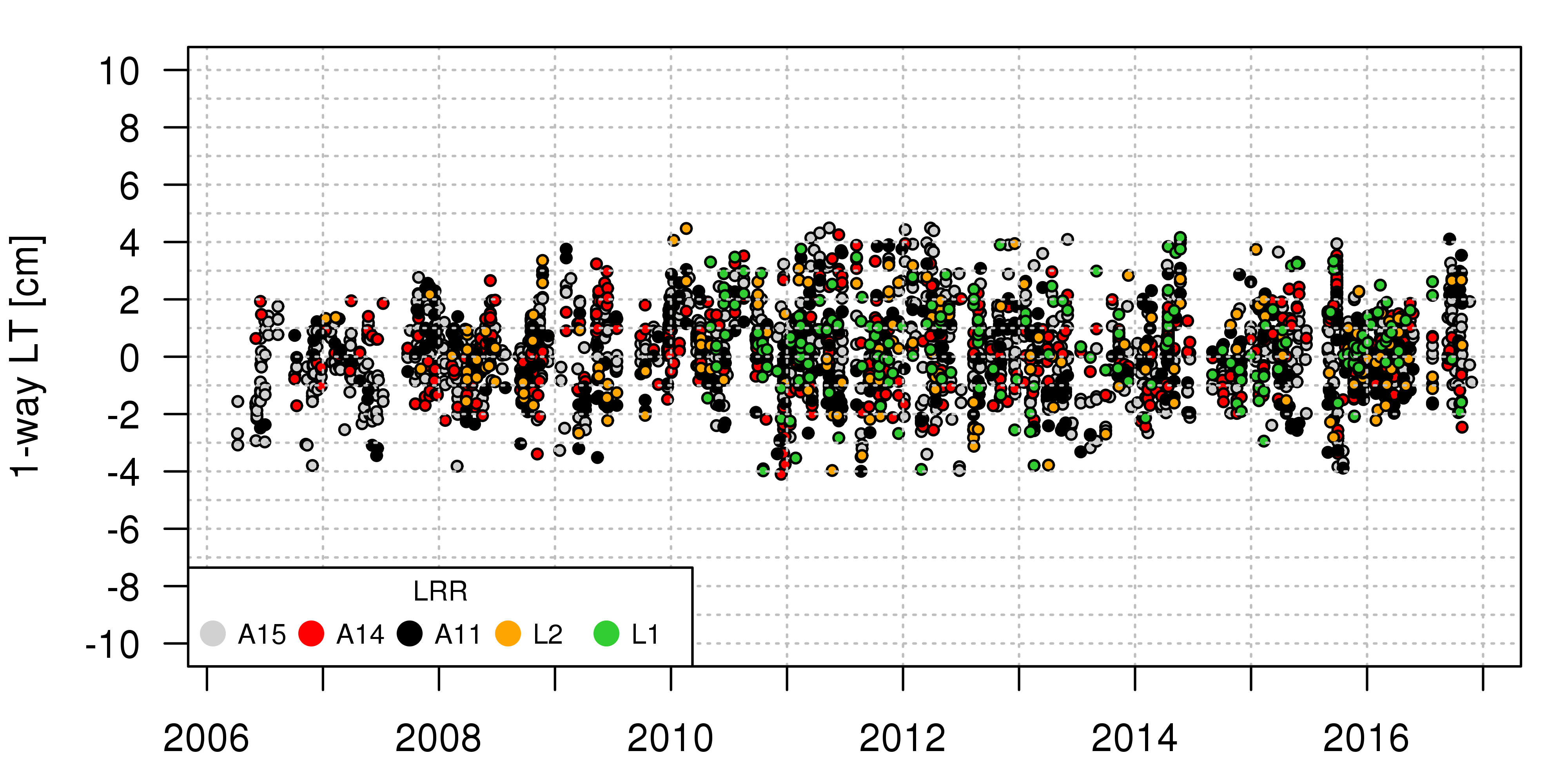}
    \caption{Post-fit residuals in (cm) vs time (year) obtained with INPOP$_{\textrm{G+IR}}$ specification (sec. \ref{appendix_est}) for APOLLO station}
    \label{fig:pf_Apollo}
    \end{figure}

\begin{table*}
    \caption{Grasse Reflector-wise statistics computed using post-fit residuals obtained with INPOP$_{\textrm{G}}$ and INPOP$_{\textrm{G+IR}}$, within the fit intervals 01/01/2015 to 01/01/2017 (with a 3-$\sigma$ filter), with the WRMS in m (RMS weighted by number of observation from each reflector). }
    \label{A15_bias_stats_1}
    \begin{tabular}{|c|cccc }
    \hline 
    \multicolumn{5}{c|}{\textbf{Grasse} }                                \\ \hline
    \textbf{LRRR}     & \textbf{INPOP$_{\textrm{G}}$}   & \textbf{INPOP$_{\textrm{G+IR}}$}   & \textbf{\% change}   & \textbf{NPTs}    \\ \hline
    A15               & 0.0183                & 0.0181                & 1.1     & 1018          \\  
    A14               & 0.0203                & 0.0177                & 12.8     &172          \\
    A11               & 0.0267                & 0.0239                & 10.5     &215           \\ 
    L1                & 0.0215                & 0.0166         &\textbf{22.8}    &265              \\ 
    L2                & 0.0246                & 0.0215                & 12.6     & 256          \\ \hline
    \textbf{WRMS}     & 0.0207    & 0.0189        & 9.5  & 1926        \\ \hline   
    \end{tabular}
    \end{table*}

    \begin{table*}
    \caption{APOLLO Reflector-wise statistics computed using post-fit residuals obtained with INPOP$_{\textrm{G}}$ and INPOP$_{\textrm{G+IR}}$, within the fit intervals 01/01/2015 to 01/01/2017 (with a 3-$\sigma$ filter), with the WRMS in m (RMS weighted by number of observation from each reflector). }
    \label{A15_bias_stats_2}
    \begin{tabular}{|c|cccc }
    \hline 
                        \multicolumn{5}{c|}{\textbf{APOLLO} }\\ \hline
    \textbf{LRRR}     &\textbf{INPOP$_{\textrm{G}}$}   &\textbf{INPOP$_{\textrm{G+IR}}$}   & \textbf{\% change}   & \textbf{NPTs} \\ \hline
    A15              &  0.0127                & 0.0127        & 0.2       & 344 \\  
    A14              &  0.0192                & 0.0177        & 7.8       &176 \\
    A11              &  0.0185                & 0.0169        & 8.7       &164 \\ 
    L1               &  0.0186                & 0.0157    &\textbf{15.6}  &89  \\ 
    L2               &  0.0136                & 0.0137        & -0.7      &64  \\ \hline
    \textbf{WRMS}    &  0.0159                & 0.0149  &6.7  & 837 \\ \hline   
    \end{tabular}
    \end{table*}

  \subsection{Results}
  \label{appendix_est}
  Table (\ref{est_extMoon}) gives the list of the adjusted parameters related to the lunar interior when Table (\ref{fixed_param}) provides a list of the fixed parameters. The fitted coordinates of the Moon reflectors and of the LLR stations can be found in \cite{Viswanathan2017a}. As the LLR observations are not included in the construction of the ITRF \cite[]{2016JGRE..121.6109L}, small corrections to the LLR station coordinates help for the improvement of LLR residuals during the construction of the lunar ephemerides. The Earth Orientation parameters (EOP) and the modeling of the Earth rotation are however kept fixed to the IERS convention (see section \ref{appendixA}).

  The solution INPOP$_{\textrm{G}}$ with an axis-symmetric core fitted to LLR observations serves as a validation of our lunar model and analysis procedure, against the DE430 JPL planetary and lunar ephemeris analysis described in \cite{Folkner2014} and EPM IAA RAS ephemeris in \cite{Pavlov2016}. Only 532 nm wavelength LLR data are used for matching with the DE430 and EPM ephemeris. 
  In \cite{Folkner2014,Pavlov2016} and INPOP$_{\textrm{G}}$, gravity field coefficients up-to degree and order 6 are used for the Moon (GL0660b from \cite{Konopliv2013}) and the Earth (GGM05C from \cite{Ries2016} for INPOP17a ephemeris and EGM2008 from \cite{Pavlis2012,Pavlis2013} for DE/EPM ephemerides). Coefficients $C_{32}$, $S_{32}$ and $C_{33}$ are then included in the fit parameters as they improve the overall post-fit residuals. 
  For INPOP$_{\textrm{G}}$, the improvement of the formal uncertainty compared to \cite{Pavlov2016}, especially in the estimation of parameter $k_v/C_T$ indicates a strong dissipation mechanism within the Moon, through viscous torques at the fluid core-mantle boundary.
  
  Differences between GL0660b values and fitted $C_{32}$, $S_{32}$ and $C_{33}$ from \cite{Folkner2014}, \cite{Pavlov2016} or in INPOP$_{\textrm{G}}$, are several orders of magnitude greater than the mean GRAIL uncertainties (see \cite{Konopliv2013}). These results suggest that some significant effects impacting the LLR observations, are absorbed by the adjustment of the degree-3 of the full Moon gravity field.
  
  The solution INPOP$_{\textrm{G+IR}}$ refers to the addition of two years of IR LLR observations \cite[]{Courde2017} described in section (\ref{npt}) and built in following the same specification as of INPOP$_{\textrm{G}}$. 
 
  This dataset is weighted at the same level as the APOLLO station normal points within the estimation procedure (see section \ref{appendixB}).

  The first outcome from the introduction of the IR data sets is the improvement of the post-fit residuals obtained for L1 reflector as one can see on Tables (\ref{A15_bias_stats_1} and \ref{A15_bias_stats_2}) and on Figures (\ref{fig:pf_rest} to \ref{fig:pf_Apollo}). This is due to the increase of normal points obtained for this reflector as discussed in section (\ref{sel_bias}). 

  The second conclusion is that because of only two years on data, the improvement brought by the addition of IR data on the estimated parameters characterizing the Moon and its inner structure is significant, especially for those quantifying the dissipation mechanism such as $Q_{27.212}$ and $\tau_{M}$ with a decreasing uncertainty or $\frac{k_{v}}{C_{T}}$ and $f_c$ with a significant change in the fitted value (see Table \ref{est_extMoon}).

A significant global improvement is  noticeable when one compares post-fit residuals obtained with INPOP$_{\textrm{G}}$ and with INPOP$_{\textrm{G+IR}}$ with those obtained with INPOP13c as presented in \cite{Fienga2014} or in Tables \ref{A15_bias_stats_1} and \ref{A15_bias_stats_2}.
  Finally one should notice in Table (\ref{NPT_span}) the 1.15 cm obtained for the post-fit weighted RMS obtained for the 3 years of the last period of the APOLLO data (group D) as well as that for the IR Grasse station. 

\section{Test of the equivalence principle}
\label{fundphy}
\subsection{Context}
Among all possibilities to test General Relativity (GR), the tests of the motion of massive bodies as well as the propagation of light in the solar system, were historically the first ones, and still provide the highest accuracies for several aspects of gravity tests (see \cite{joyce2015,berti2015,Yunes2016} for recent overviews of constraints on alternative theories from many different types of observations). This is in part due to the fact that the dynamics of the solar system is well understood and supported by a long history of observational data.  

In GR, not only do test particles with different compositions fall equally in a given gravitational field, but also extended bodies with different gravitational self-energies. While a deviation from the former case would indicate a violation of the Weak Equivalence Principle (WEP), a deviation from the latter case would be a sign of a violation of the Strong Equivalence Principle (SEP) (\cite{will:2014lr}). Violations of the Equivalence Principles are predicted by a number of modifications of GR, often intending to suggest a solution for the problems of Dark Energy and Dark Matter \cite{capozziello2011, joyce2015,berti2015} and/or to put gravity in the context of Quantum Field Theory \cite{kostelecky2004,woodard2009,donoghue2017}. The Universality of Free Fall (UFF), an important part of the Equivalence Principle, is currently tested at a level of about 10$^{-13}$ with torsion balances \cite[]{2003PhRvD..68f2002A} and LLR analyses \cite[]{Williams2012}. 

As the Earth and the Moon both fall in the gravitational field of the Sun --- and because they neither have the same compositions, nor the same gravitational self-energies --- the Earth-Moon system is an ideal probe of both the WEP and the SEP, while torsion balance \cite[]{2003PhRvD..68f2002A} or MICROSCOPE \cite[]{2014AdSpR..54.1119L} are only sensitive to violations of the WEP.

In this paper, we implemented the equations given in \cite{Williams2012} and introduce in the INPOP fit, the differences between the accelerations of the Moon and the Earth.

The aim of this work is first to give the most general constraint in terms of acceleration differences without assuming metric theories or other types of alternative theories (section \ref{resultsep}). In a second step (section \ref{thiner}), we propose two interpretations : one following the usual formalism proposed by Nordtvedt (see, e.g., \cite[]{nordtvedt:2014sc} and references therein), and the other following the dilaton theory \cite[]{damour:1994np,hees:2015ax,minazzoli:2016pr}.

\subsection{Method}

In order to test possible violations of GR in terms of UFF, a supplementary acceleration is introduced in the geocentric equation of motion of the Moon, such that the UFF violation related difference between the Moon and the Earth accelerations reads \cite[]{1968PhRv..170.1186N}:
\begin{eqnarray}
\Delta \bm a^{\overline{UFF}}\equiv (\bm a_M - \bm a_E)^{\overline{UFF}} = \bm a_{E} \Delta_{ESM}
\label{UFF}
\end{eqnarray}
$\Delta_{ESM}$ is estimated in the LLR adjustment together with the other parameters of the lunar ephemerides given in Table (\ref{est_extMoon}). In what follows, we shall name $\Delta_{ESM}$ ``UFF violation parameter''. $ESM$ stands for the three bodies involved, namely the Earth, the Sun and the Moon respectively. As we shall see in Sec. \ref{dil_theory}, some theoretical models induce a dependence of the UFF violation parameter on the composition of the Sun, in addition to the ``more usual'' dependence on the compositions and on the gravitational binding energies of the Moon and the Earth. 

In order to estimate $\Delta_{ESM}$ with the appropriate accuracy, one should correct for supplementary effects such as the solar radiation pressure and thermal expansion of the retro-reflectors \cite[]{Vokrouhlicky1997,Williams2012}. An empirical correction on the radial perturbation ($\Delta r_{EM}$) induced by the UFF test has to be applied. 
For instance, with some simplifying approximations (\cite{nordtvedt:2014sc}), one can show that the UFF additional acceleration would indeed lead to an additional radial perturbation ($\Delta r_{EM}$) of the Moon's orbit towards the direction of the Sun given by:
\begin{eqnarray}
\Delta r_{EM}= S \Delta_{ESM} \cos D,
\label{SRP}
\end{eqnarray}
where $S$ is a scaling factor of about $-3 \times 10^{10}$ m \cite[]{Williams2012} and $D$ is the synodic angle. A correction ${\Delta r = 3.0 \pm 0.5}$ mm \cite[]{Vokrouhlicky1997,Williams2012} is then applied in order to correct for solar radiation pressure and thermal radiation of the retro-reflectors, and a new corrected value of $\Delta_{ESM}$ is then deduced (see Table \ref{mgmi}).

\subsection{Results}
\label{resultsep}

Fits were performed including in addition to the previous fitted parameters presented in Table (\ref{est_extMoon}), the UFF violation parameter $\Delta_{ESM}$ given in Eqn. (\ref{UFF}). Two different fits were considered including 532 nm and 1064 nm data sets (solution labeled INPOP$_{\textrm{G+IR}}$), or just the 532 nm data sets (solution labeled INPOP$_{\textrm{G}}$). A supplementary adjustment was also performed for a better comparison to the previous determination from other LLR analysis groups, which were limited to a data sample up to 2011 (labeled as limited data). Results are given in Table (\ref{mgmi}).

The additional acceleration of the Moon orbit in the direction of the Sun correlates with a coefficient of 0.95 and 0.90 with GM$_{\textrm{EMB}}$ and the Earth-Moon mass ratio (EMRAT), respectively. In all the solutions w.r.t LLR EP estimation, the gravitational mass of the Earth Moon barycenter (GM$_{\textrm{EMB}}$) remains as a fit parameter due its high correlation with the EP parameter ($\Delta_{ESM}$). EMRAT was estimated from a joint planetary solution and kept fixed during LLR EP tests (for all INPOP solutions in Table \ref{mgmi}) due to its weak determination from LLR. 

A test solution that fitted EMRAT, with GM$_{\textrm{EMB}}$ as a fixed parameter, gives an estimate of $\Delta_{ESM} = (8 \pm 7.0 ) \times10^{-14}$. However, the value of EMRAT estimated from an LLR only solution has an uncertainty of one order of magnitude greater than that obtained from the joint planetary fit. This is also consistent with a similar result by \cite{Williams2009}. As a result, EMRAT was not included as a fit parameter for the estimates provided in Table (\ref{mgmi}), as it resulted in a degraded fit of the overall solution.

\cite{Williams2012} show that including annual nutation components of the Earth pole direction in space, to the list of fitted parameters during the estimation of LLR EP solution, increases the uncertainty of the estimated UFF violation parameter ($\Delta_{ESM}$) by 2.5 times. Moreover, it is to be noted that within Table (\ref{mgmi}), the solutions by \cite{Williams2009,Williams2012,Muller2012} use the IERS2003 \cite[]{McCarthy2004} recommendations within the reduction model, while all INPOP17 solutions use IERS 2010 \cite[]{IERS2010} recommendations. The notable difference between the two IERS models impacting the LLR EP estimation is expected to be from the precession-nutation of the celestial intermediate pole (CIP) within the ITRS-GCRS transformation \citet[p.~8]{IERS2010}.

Eqn. (\ref{SRP}) shows the dependence of $\Delta_{ESM}$ w.r.t the cosine of the lunar orbit synodic angle, synonymous with the illumination cycle of the lunar phases. Due to the difficulties involved with ranging to the Moon during the lunar phases with the maximum value of $\cos{D}$ (New and Full Moon) as described in section (\ref{phas_bias}), the LLR observations during these phases remain scarce. The availability of IR LLR observations from Grasse, contributes to the improvement of this situation, as shown in Fig. (\ref{NPT_syn_hist}). This is reflected in the improvement of the uncertainty of the estimated value of $\Delta_{ESM}$ by 14 \%, with solutions including the IR LLR data. 

Using both IR and green wavelength data, and empirically correcting for the radial perturbation for effects related to solar radiation pressure and thermal expansion, our final result on the UFF violation parameter is given by (see, also, Table \ref{mgmi})
\begin{eqnarray}
\Delta_{ESM} = (-3.8 \pm 7.1) \times10^{-14}
\end{eqnarray}

The continuation of the IR observational sessions at Grasse will help to continue the improvement in the $\Delta_{ESM}$ estimations.

An observable bias in the differential radial perturbation of the lunar orbit w.r.t the Earth, towards the direction of the Sun, if significant and not accounted for within the dynamical model, would result in a false indication of the violation of the principle of equivalence estimated with the LLR observations. \cite{Oberst2012} show the distribution of meteoroid impacts with the lunar phase. Peaks within the histogram in \citet[p~186]{Oberst2012} indicate a non-uniform temporal distribution with a non-negligible increase in both small and large impacts during the New and Full Moon phase. Future improvements to the LLR EP estimation must consider the impact of such a bias that could potentially be absorbed during the fit by the LLR UFF violation parameter $\Delta_{ESM}$.

\subsection{Theoretical interpretations}
\label{thiner}

\subsubsection{Nordtvedt's interpretation: gravitational versus inertial masses}
\label{classic}
Although equations of motion are developed at the post-Newtonian level in INPOP \cite[]{Moyer2003}, violations of the UFF can be cast entirely in the Newtonian equation of motion with sufficient accuracy. As described by Nordtvedt \cite[]{1968PhRv..170.1186N}, a difference of the inertial ($m^I$) and gravitational ($m^G$) masses would lead to an alteration of body trajectories in celestial mechanics according to the following equation:
\begin{eqnarray}
\bm {a}_T=-\left(\frac{m^G}{m^I} \right)_T \sum_{A\neq T} \frac{G m^G_A}{r_{AT}^3}\bm r_{AT} \label{eq:nordt_motion},
\end{eqnarray}
where $\bm r_{AT}= \bm x_T - \bm x_A$ and $G$ is the constant of Newton.

Following \cite{Williams2012},  the relative acceleration at the Newtonian level between the Earth and the Moon reads
\begin{eqnarray}
\bm a_M - \bm a_E= -  \frac{ G \mu}{r_{EM}^3}\bm r_{EM}+   G m^G_S\left[ \frac{\bm r_{SE}}{r_{SE}^3}-\frac{\bm r_{SM}}{r_{SM}^3} \right] + \nonumber \\
+   G  m^G_S\left[ \frac{\bm r_{SE}}{r_{SE}^3} \left(\left(\frac{m^G}{m^I} \right)_E -1\right)- \frac{\bm r_{SM}}{r_{SM}^3} \left(\left(\frac{m^G}{m^I} \right)_M -1\right) \right], \label{eq:williams}
\end{eqnarray}
with $\mu\equiv  m^G_M +  m^G_E+ \left(\left(\frac{m^G}{m^I} \right)_E -1\right)m^G_M+\left(\left(\frac{m^G}{m^I} \right)_M -1\right) m^G_E$. $\left(\frac{m^G}{m^I} \right)_E$ and $\left(\frac{m^G}{m^I} \right)_M$ are the ratios between the gravitational and the inertial masses of the Earth and Moon respectively.

With ephemeris, the first term of Eqn. (\ref{eq:EMacc}) does not lead to a sensitive test of the UFF, because it is absorbed in the fit of the parameter $m^G_M+m^G_E$ \citep[e.g.]{Williams2012}. The last term, on the other side, does. At leading order, one can approximate both distances appearing in this last term as being approximately equal. One gets
\begin{eqnarray}
\Delta \bm a^{\overline{UFF}} &\equiv& (\bm a_M - \bm a_E)^{\overline{UFF}} \nonumber \\
&\approx&  G m^G_S\left[ \frac{\bm r_{SE}}{r_{SE}^3} \left(\left(\frac{m^G}{m^I} \right)_E -1\right)- \frac{\bm r_{SM}}{r_{SM}^3} \left(\left(\frac{m^G}{m^I} \right)_M -1\right) \right]\nonumber \\
&\approx&  \bm a_{E}\left[ \left(\left(\frac{m^G}{m^I} \right)_E -1\right)-  \left(\left(\frac{m^G}{m^I} \right)_M -1\right) \right] \nonumber \\
&\equiv& \bm a_{E} \Delta_{ESM}
\end{eqnarray}
with
\begin{eqnarray}
\Delta_{ESM} =\left[\left(\frac{m^G}{m^I} \right)_E-  \left(\frac{m^G}{m^I} \right)_M \right].\label{eq:Delta_usual}
\end{eqnarray}
One recovers Eqn. (\ref{UFF}). Therefore, in this context, constraints on $\Delta_{ESM}$ can be interpreted as constraints on the difference of the gravitational to inertial mass ratios between the Earth and the Moon. 

Furthermore, the LLR test of UFF captures a combined effect of the SEP, from the differences in the gravitational self-energies, and the WEP due to compositional differences, of the Earth-Moon system. In general, one has:
\begin{eqnarray}
\label{sep_wep}
\Delta_{ESM} =\Delta_{ESM}^{WEP} + \Delta_{ESM}^{SEP}
\end{eqnarray}
In order to separate the effects of WEP, we rely on results from laboratory experiments that simulate the composition of the core and the mantle materials of the Earth-Moon system. One such estimate is provided by \cite{Adelberger2001}, that translates to the following mass ratios difference:
 \begin{eqnarray}
  \label{WEP_Adelberge}
 \Delta_{ESM}^{WEP}&=&  \Big[\Big(\frac{m^G}{m^I}\Big)_{E} - \Big(\frac{m^G}{m^I}\Big)_{M}\Big]_{WEP} \\ &=& (1.0 \pm 1.4) \times 10^{-13} 
  \end{eqnarray}
The results of the estimation of the derived value of the SEP from LLR is provided in Table (\ref{sep_mgmi}). From the values given in Table (\ref{sep_mgmi}) it is also possible to deduce the Nordtvedt parameter ($\eta$) defined as:
\begin{eqnarray}
  \label{eta_sep_1}
  \Delta_{ESM}^{SEP} &=& \eta_{SEP} \left[\left(\frac{|\Omega|}{m~c^2}\right)_E - \left(\frac{|\Omega|}{m~c^2}\right)_M\right] \\ &\approx&  \eta_{SEP} \times (-4.45 \times 10^{-10})
  \end{eqnarray}
where $\Omega$ and $m c^2$ are the gravitational binding and rest mass energies respectively for the Earth and the Moon (subscripts E and M respectively). The value of $-4.45 \times 10^{-10}$ is obtained from \citet[Eqn. 7]{Williams2009}.

However, all metric theories lead to a violation of the SEP only. Therefore, for metric theories, it is irrelevant to try to separate violation effects of the WEP and SEP, as the WEP is intrinsically respected. The estimates of $\eta_{SEP}^\mathrm{metric}$ in such cases are provided in Table (\ref{mgmi}).

\subsubsection{Dilaton theory and a generalization of the Nordtvedt interpretation}
\label{dil_theory}
Starting from a general dilaton theory,  a more general equation governing celestial mechanics than (\ref{eq:nordt_motion}) has been found to be \cite[]{hees:2015ax,minazzoli:2016pr}
\begin{eqnarray}
\bm {a}_T=-\sum_{A\neq T} \frac{G m^G_A}{r_{AT}^3}\bm r_{AT}\left(1+\delta_T+\delta_{AT}\right) \label{eq:EIH_s},
\end{eqnarray}
 The coefficients $\delta_T$ and $\delta_{AT}$ parametrize the violation of the UFF.  In this expression the inertial mass $m^I_A$ writes in terms of the gravitational mass $m^G_A$ as $m^G_A=(1+\delta_A) m^I_A$ \cite[]{hees:2015ax,minazzoli:2016pr}. Of course, since $m_A^G/m_A^I=1+\delta_A$, one recovers Eqn. (\ref{eq:nordt_motion}) when $\delta_{AB}=0$ for all $A$ and $B$. From Eqn. (\ref{eq:EIH_s}), one can check that the gravitational force in this context still satisfies Newton's third law of motion:
\begin{equation}
m^I_A\bm {a}_A =\frac{G m^I_A m^I_B}{r_{AB}^3}\bm r_{AB}\left(1+\delta_A+\delta_B+\delta_{AB}\right)=- m^I_B\bm {a}_B.
\end{equation}

In the dilaton theory, the $\delta$ coefficients are functions of ``dilatonic charges'' and of the fundamental parameters of the theory \cite[]{Damour2010,hees:2015ax,minazzoli:2016pr}. However, in what follows, we will consider the phenomenology based on the $\delta$ parameters independently of its theoretical origin, as a similar phenomenology may occur in a different theoretical framework.

In general, $\delta_T$ can be decomposed into two contributions: one from a violation of the WEP and one from a violation of the SEP:
\begin{equation}
\label{eta_sep}
\delta_T=\delta_T^{WEP}+\delta_T^{SEP},\qquad \textrm{with}\qquad \delta_T^{SEP} = \eta \frac{|\Omega_T|}{m_T~c^2},
\end{equation}
The quantity $\delta_T^{SEP}$ depends only on the gravitational energy content of the body $T$. On the other hand, $\delta_T^{WEP}$ depends on the composition of the falling body $T$ (\cite{Damour2010,hees:2015ax,minazzoli:2016pr}). In some theoretical situations (see e.g. \cite{Damour2010}), if $\delta_T^{WEP} \neq 0$, then $\delta_T^{WEP} \gg  \delta_T^{SEP}$, such that one can have either a clean WEP violation, or a clean SEP violation.

Like the parameter $\delta_T^{WEP}$, $\delta_{AT}$ depends on the composition of the falling bodies. However, unlike $\delta_T^{WEP}$, it also depends on the composition of the body $A$ that is source of the gravitational field in which the body $T$ is falling (\cite{hees:2015ax,minazzoli:2016pr}). As a consequence, the relative acceleration of two test particles with different composition cannot only be related to the ratios between their gravitational to inertial masses in general (i.e. $m_A^G/m_A^I=1+\delta_A$). This contrasts with the usual interpretation (see for instance (\cite{Williams2012})). However, with some theoretical models, $\delta_T^{WEP} \gg \delta_{AT}$ (\cite{Damour2010,hees:2015ax,minazzoli:2016pr}).

At the Newtonian level, the relative acceleration between the Earth and the Moon reads
\begin{eqnarray}
\bm a_M - \bm a_E = -  \frac{ G \mu}{r_{EM}^3}\bm r_{EM}+   G m^G_S\left[ \frac{\bm r_{SE}}{r_{SE}^3}-\frac{\bm r_{SM}}{r_{SM}^3} \right] \nonumber \\
+   G  m^G_S\left[ \frac{\bm r_{SE}}{r_{SE}^3} (\delta_E+\delta_{SE})- \frac{\bm r_{SM}}{r_{SM}^3} (\delta_M+\delta_{SM}) \right], \label{eq:EMacc}
\end{eqnarray}
with $\mu\equiv  m^G_M +  m^G_E+ (\delta_E+\delta_{EM})m^G_M+(\delta_M+\delta_{EM}) m^G_E$. 
As discussed already in the previous subsection, the first term of Eqn. (\ref{eq:EMacc}) does not lead to a sensitive test of the UFF, because it can be absorbed in the fit of the parameter $m^G_M+m^G_E$ \citep[e.g.][]{Williams2012}. The last term, on the other side, does. At leading order, one can approximate both distances appearing in this last term as being approximately equal. One therefore has
\begin{eqnarray}
\Delta \bm a^{\overline{UFF}} &\equiv& (\bm a_M - \bm a_E)^{\overline{UFF}} \nonumber \\
&\approx&  G m^G_S\left[ \frac{\bm r_{SE}}{r_{SE}^3} (\delta_E+\delta_{SE})- \frac{\bm r_{SM}}{r_{SM}^3} (\delta_M+\delta_{SM}) \right] \nonumber \\
&\approx& \bm a_{E}  \left[(\delta_E+\delta_{SE})- (\delta_M+\delta_{SM})\right] \nonumber\\
&\equiv& \bm a_{E} \Delta_{ESM}
\end{eqnarray}
where  $\Delta \bm a^{\overline{UFF}}$ is the part of the relative acceleration between the Earth and the Moon that violates the UFF. Once again, one recovers Eqn. (\ref{UFF}) --- although its theoretical interpretation is different compared to the previous subsection.

When $\delta_{SM}=\delta_{SE}$, and especially when $\delta_{SM}=\delta_{SE}=0$, one recovers the usual Eqn. (\ref{eq:Delta_usual}). But it is not the case in general because the composition of the Sun may affect the dynamics in some cases as well. Therefore, in a more general context than in section (\ref{classic}), constraints on $\Delta_{ESM}$ cannot be uniquely interpreted as constraints on the difference of the gravitational to inertial mass ratios between the Earth and the Moon. 

As a consequence, from a pure phenomenological point of view --- or, equivalently, from an agnostic point of view --- one shouldn't interpret $\Delta_{ESM}$ in terms of gravitational to inertial mass ratios only. Indeed, a more general expression of the UFF violating parameter is given by
\begin{eqnarray}
\Delta_{ESM} =\left[(\delta_E+\delta_{SE})- (\delta_M+\delta_{SM}) \right],\label{eq:Delta_dilaton}
\end{eqnarray}
where one can see that the Sun's composition may affect the dynamics as well, through the coefficients $\delta_{SE}$ and $\delta_{SM}$. 

(Otherwise, see a discussion on how to decorrelate the dilaton parameters from planetary ephemeris in \cite[]{2017arXiv170505244M}).

\section{Conclusions and future work}
In this paper, we present an improvement in the lunar dynamical model of INPOP ephemeris (version 17a) compared to the previous release (version 13c). The model is fitted to the LLR observations between 1969-2017, following the model recommendations from IERS 2010 \cite[]{IERS2010}. The lunar parameter estimates obtained with the new solution are provided in Table (\ref{est_extMoon}) with comparisons to that obtained by other LLR analyses groups. The improvement brought by the new IR LLR data from Grasse station on the parameter estimates is characterized. The post-fit LLR residuals obtained with INPOP17a are between 1.15 cm to 1.95 cm over 10 years of APOLLO data and 1.47 cm over 2 years of the new IR LLR data from Grasse \cite[]{Viswanathan2017a}.
Our solution benefits also of the better spatial and temporal distribution of the IR Grasse data with an improvement of 14$\%$ of the UFF tests and better estimations of the Moon dissipation parameters.

We take advantage of the lunar ephemeris improvements to perform new tests of the universality of free fall. A general constraint is obtained using INPOP, in terms of the differences in the acceleration of the Earth and the Moon towards the Sun. In addition to the Nordtvedt interpretation of \cite{1968PhRv..170.1186N} (provided in section \ref{classic}), we propose an alternative interpretation and a generalization of the usual interpretation from the point of view of the dilation theory \cite[]{damour:1994np,hees:2015ax,minazzoli:2016pr} (provided in section \ref{dil_theory}). We obtain an estimate of the UFF violating parameter $\Delta_{ESM} = (-3.8 \pm 7.1) \times10^{-14}$, showing no violation of the principle of equivalence at the level of $10^{-14}$.

Thermal expansion of the retro-reflectors and solar radiation pressure are currently employed as empirical corrections following \cite{Vokrouhlicky1997,Williams2009}. Future LLR analysis will consider an implementation of these effects within the reduction procedure, so as to improve the uncertainty of the EP test. \cite{Oberst2012} show the distribution of meteoroid impacts with the lunar phase, indicating a non-uniform temporal distribution during the New and Full Moon phase which could impact the test of EP. The impact of this effect needs to be characterized during the EP test, to be considered as negligible at the present LLR accuracy.


The use of a strictly GRAIL-derived gravity field model \cite[]{Konopliv2013} highlights longitude libration signatures well above the LLR noise floor, arising from unmodeled effects in lunar ephemeris \cite[]{Viswanathan2017b}. Other LLR analyses groups \cite[]{Folkner2009,Folkner2014,Pavlov2016} prefer to fit the degree-3 components away from GRAIL-derived gravity field coefficients. A work is in progress to identify the cause of the low-degree spacecraft-derived gravity field inconsistency.

\section*{Acknowledgments}
The authors extend their sincere gratitude to all the observers and engineers at Grasse, APOLLO, McDonald, Matera and Haleakala LLR stations for providing timely and accurate observations over the past 48 years.

 \begin{table*}
    \label{fixed_param}
    \caption{Fixed parameters for the Earth-Moon system.}
    \begin{tabular}{llccc}
    \hline
   \textbf{Parameter}                                             & \textbf{Units}  & \textbf{INPOP}  & \textbf{DE430}  & \textbf{EPM}  \\ \midrule
           $(EMRAT^{\dag} - 81.300570)\times 10^{6}$              &                 & 1.87          & -0.92              & -0.92            \\ 
           $(R_E    -   6378.1366) \times 10^{4}$                 & km              & 0.0           & -3              & 0.0           \\          
           $(\dot{J}_{2E} -2.6\times 10^{-11}) $                  & year$^{-1}$     & 0.0           & 0.0             & 0.0           \\ 
           $(k_{20,E} - 0.335)$                                   &                 & 0.0           & 0.0             & 0.0           \\ 
           $(k_{21,E} - 0.32)$                                    &                 & 0.0           & 0.0             & 0.0           \\ 
           $(k_{22,E} - 0.30102)$                                 &                 & -0.01902      & 0.01898         & -0.01902      \\ 
           $(\tau_{O0,E} - 7.8\times 10^{-2}) \times 10^{2} $     & day             & 0.0           & -1.4            & 0.0           \\ 
           $(\tau_{O1,E} +4.4\times 10^{-2}) $                    & day             & 0.0           & 0.0$^{\ddag}$   & 0.0           \\ 
           $\tau_{O2,E} +1.13\times 10^{-1})\times 10^{1}  $      & day             & 0.0           & 0.13            & 0.0           \\ 
           $(R_M - 1738.0)$                                       & km              & 0.0           & 0.0             & 0.0           \\          
           $(\alpha_{C} - 7.0\times 10^{-4}) $                    &                 & 0.0           & 0.0             & 0.0           \\ 
           $ (k_{2,M} -0.024059)$                                 &                 & 0.0           & 0.0             & 0.0           \\ 
           $ (l_{2} - 0.0107)$                                    &                 & 0.0           & 0.0             & 0.0           \\ 
           \hline
           \multicolumn{5}{l}{$^\dag$: EMRAT is fitted during the joint analysis between the lunar and planetary part.} \\
           \multicolumn{5}{l}{$^\ddag$: $\tau_{O1,E}$ in \cite{Folkner2014} given as -0.0044 is a typographical error.}
    \end{tabular}
   \end{table*}

\begin{table*}
    \label{est_extMoon}
    \caption{Extended body parameters for the Earth and the Moon. Uncertainties for INPOP$_{\textrm{G}}$  and INPOP$_{\textrm{G+IR}}$ (1-$\sigma$) are obtained from a 5 \% jackknife (JK), while other solutions (DE430 and EPM) are assumed as (1-$\sigma$) formal uncertainties. $^\dag$: $C_{32}$, $S_{32}$ and $C_{33}$ are reference values from the GRAIL analysis by \protect\cite{Konopliv2013}. $^\ddag$: $h_2$ reference value from LRO-LOLA analysis by \protect\cite{Mazarico2014}. $^{*}$ : derived quantity}
    \begin{tabular}{llcccc}
           \hline
           \textbf{Parameter}                                       & \textbf{Units}    & \textbf{INPOP$_{\textrm{G}}$}     & \textbf{INPOP$_{\textrm{G+IR}}$} & \textbf{DE430}    & \textbf{EPM}    \\ \midrule 
           $(GM_{\textrm{EMB}}-8.997011400 \times 10^{-10}) \times 10^{19} $ & AU$^3$/day$^2$    & $4 \pm 2$                         & 4 $\pm$ 2                        & -10               & $10 \pm 5$      \\ 
           $(\tau_{R1,E} - 7.3\times 10^{-3}) \times 10^{5}$        & day               & $0 \pm 4$                         & 6 $\pm$ 3                        & $6 \pm 30$        & $57 \pm 5$      \\ 
           $(\tau_{R2,E}-2.8\times 10^{-3}) \times 10^{5}$          & day               & $9.2 \pm 0.4$                     & 8.7 $\pm$ 0.3                    & $-27 \pm 2$       & $5.5 \pm 0.4$   \\ 
           $(C_{T}/(m_{M}R^2)- 0.393140) \times 10^{6}$             &                   & $6.9 \pm 0.2$                     & 8.2 $\pm$ 0.2                    & $2^{*} $          & $2^{*} $  \\ 
           $(C_{32} - 4.8404981\times 10^{-6 \dag}) \times 10^{9}$  &                   & $4.1 \pm 0.3$                     & 3.9 $\pm$ 0.3                    & $4.4 $            & $4.4\pm 0.1$    \\ 
           $(S_{32} - 1.6661414\times 10^{-6 \dag}) \times 10^{8}$  &                   & $1.707 \pm 0.006$                 & 1.666 $\pm$ 0.006                & $1.84 $           & $1.84\pm 0.02$  \\ 
           $(C_{33} - 1.7116596\times 10^{-6 \dag}) \times 10^{8}$  &                   & $-1.19 \pm 0.04$                  & -2.40 $\pm$ 0.04                 & $-3.6 $           & $-4.2 \pm 0.2$  \\ 
           $(\tau_{M} - 9\times 10^{-2}) \times 10^{4}$             & day               & $-14 \pm 5$                       & -35 $\pm$ 3                      & $58.0 \pm 100$    & $60 \pm 10$     \\ 
           $(\frac{k_{v}}{C_{T}}-1.6\times 10^{-8}) \times 10^{10}$ & day$^{-1}$        & $12.7 \pm 0.4$                    & 15.3 $\pm$ 0.5                   & $4.0 \pm 10.0$    & $3.0 \pm 2.0$   \\ 
           $(f_c-2.1\times 10^{-4}) \times 10^{6}$                  &                   & $37 \pm 3$                        & 42 $\pm$ 3                       & $36 \pm 28$       & $37\pm 4$       \\ 
           $(h_{2}- 3.71 \times 10^{-2 \ddag}) \times 10^{3}$       &                   & $6.3\pm 0.2$                      & 6.8$\pm$ 0.2                     & $11.0\pm 6 $      & $6 \pm 1$       \\
           $Q_{27.212} - 45$ (derived)                              &                   & $3.9 \pm 0.5$                     & 5.0 $\pm$ 0.2                    & $0 \pm 5$         & $0 \pm 1$       \\ 
           \hline
    \end{tabular}
    \end{table*}

\begin{table*}
\caption{Comparison of results for the value of $\Delta_{ESM}$ (Column 4) estimated with the solution INPOP17A fitted to LLR dataset between: 1) 1969-2011 (for comparison with \protect\cite{Williams2012,Muller2012}; 2) 1969-2017 with data obtained only in Green wavelength, 3) 1969-2017 with data obtained with both Green and IR wavelength. Column 5 empirically corrects the radial perturbation from effects related to solar radiation pressure and thermal expansion of retro-reflectors using Eqn. (\ref{SRP}), with a value $\Delta{r}= 3.0 \pm 0.5$ mm \protect\cite[]{Williams2012}. Column 6 contains the value of $\Delta_{ESM}$ after applying the corrections of Column 5. Column 7 contains the Nordtvedt parameter ($\eta_{SEP}^\mathrm{metric}$) obtained using Eqn. (\ref{eta_sep_1}) with a metric theory prior.}
\label{mgmi}
\begin{tabular}{lcccccc}
\toprule
\textbf{Reference} 		             &\textbf{Data}	       & \textbf{Uncertainty}		& \textbf{estimated}      		                         &\textbf{corrected} 					& \textbf{corrected}                        & \textbf{Nordtvedt parameter$^\dag$}                       \\
				                                &\textbf{time span}	 & 		                                     & \textbf{$\Delta_{ESM}$} 	                         & \textbf{$\cos{D}$} 					& \textbf{$\Delta_{ESM}$}           & \textbf{$\eta_{SEP}^\mathrm{metric}$}                     \\ 
				                                &{[}Year{]}		       & 		                                     & [$\times10^{-14}$] 			                         & [mm] 								& [$\times10^{-14}$]                      & [$\times 10^{-4}$]                       \\ \midrule
\cite{Williams2009}$^\dag$ 		&1969-2004 		       & N/A 			                  & 3.0 $\pm$ 14.2 				                         & 2.8 $\pm$ 4.1 						& -9.6 $\pm$ 14.2                            & 2.24 $\pm$ 3.14                         \\
\cite{Williams2012} 		             &1969-2011 		       & N/A		                              & 0.3 $\pm$ 12.8 			 	                         & 2.9 $\pm$ 3.8 						& -9.9 $\pm$ 12.9                            & 2.25 $\pm$ 2.90                        \\
\cite{Muller2012}$^{\star\dag}$	&1969-2011 		       & 3-$\sigma$		                  & -14 $\pm$ 16    			                               & - 						                   & -                                                    & -                         \\ 
INPOP17A (limited data)	             &1969-2011 		       & 3-$\sigma$		                  & -3.3 $\pm$ 17.7 				                         & 4.0 $\pm$ 5.2 						& -13.5 $\pm$ 17.8                          & 3.03 $\pm$ 4.00                         \\
\cite{Hofmann2016}$^\dag$		&1969-2016 		       & 3-$\sigma$		                  & - 							                               & - 									& -3.0 $\pm$ 6.6                              & 0.67 $\pm$ 1.48                         \\
INPOP17A (Green only)	             &1969-2017 		       & 3-$\sigma$		                  & 5.2 $\pm$ 8.7 				                         & 1.5 $\pm$ 2.6 						& -5.0 $\pm$ 8.9                              & 1.12 $\pm$ 2.00                        \\
INPOP17A (Green and IR)            &1969-2017		       & 3-$\sigma$		                  & 6.4 $\pm$ 6.9 				                         & 1.1 $\pm$ 2.1 						& -3.8 $\pm$ 7.1                              & 0.85 $\pm$ 1.59                        \\ \bottomrule
\multicolumn{7}{l}{$^\star$: SRP correction not applied} \\
\multicolumn{7}{l}{$^\dag$: Thermal expansion correction not applied}\\
\multicolumn{7}{l}{$^\ddag$: derived using $\frac{|\Omega_E|}{m_E~c^2} - \frac{|\Omega_M|}{m_M~c^2}$ = -4.45$\times10^{-10}$ \protect\cite[Eqn. 6]{Williams2012}}

\end{tabular}
 \end{table*}

  \begin{table*}
  \caption{Results of SEP estimates obtained from LLR EP numerical estimates, after removing WEP component provided by laboratory experiments from \protect\cite{Adelberger2001}. The parameter $\eta_{SEP}$ is obtained from the derived SEP using Eqn. (\ref{sep_wep}) and Eqn. (\ref{eta_sep_1}). The corresponding uncertainties from the corrected value of $\Delta_{ESM}$ (Table \ref{mgmi}) and the laboratory estimate of WEP (Eqn. \ref{WEP_Adelberge}) add in quadrature for the derived value of SEP in Column 4.}
  \label{sep_mgmi}
  \begin{tabular}{lcccc}
  \toprule
  \textbf{Reference}              & \textbf{Data}                 & \textbf{Uncertainty}             & \textbf{derived SEP}                           & \textbf{Nordtvedt parameter$^\ddag$}\\
                                               & \textbf{time span}         &                                              & \textbf{$\Delta(m^G/m^I)_{ESM}$} & \textbf{$\eta_{SEP}$} \\ 
                                               &{[}Year{]}                      &                                              & [$\times10^{-13}$]                             & [$\times 10^{-4}$] \\ \midrule
  \cite{Williams2009}$^\star$           & 1969-2004                      & N/A                                      & -2.0 $\pm$ 2.0                                      & 4.4 $\pm$ 4.5   \\
  \cite{Williams2012}           & 1969-2011                      & N/A                                      & -2.0 $\pm$ 1.9                                      & 4.4 $\pm$ 4.3   \\
  \cite{Muller2012}$^\star$               & 1969-2011                      &3-$\sigma$                            & -2.4 $\pm$ 2.1                                      & 5.4 $\pm$ 4.7   \\
  INPOP17A (limited data)    & 1969-2011                     &3-$\sigma$                            & -2.4 $\pm$ 2.3                                      & 5.4 $\pm$ 5.1   \\
  INPOP17A (Green only)     & 1969-2017                     &3-$\sigma$                             & -1.5 $\pm$ 1.7                                    &  3.3 $\pm$ 3.8   \\
  INPOP17A (Green and IR) & 1969-2017                     &3-$\sigma$                             & -1.4 $\pm$ 1.6                                    &  3.1 $\pm$ 3.6  \\ 
  \bottomrule
  \multicolumn{5}{l}{$^\star$: fitted parameters include annual nutation coefficients} \\
  \multicolumn{5}{l}{$^\ddag$: derived using $\frac{|\Omega_E|}{m_E~c^2} - \frac{|\Omega_M|}{m_M~c^2}$ = -4.45$\times10^{-10}$ \protect\cite[Eqn. 6]{Williams2012}}
  \end{tabular}
  \end{table*}

\clearpage
\bibliographystyle{mnras}
\bibliography{biblio_sep} 

\begin{thebibliography}{}
\makeatletter
\relax
\def\mn@urlcharsother{\let\do\@makeother \do\$\do\&\do\#\do\^\do\_\do\%\do\~}
\def\mn@doi{\begingroup\mn@urlcharsother \@ifnextchar [ {\mn@doi@}
  {\mn@doi@[]}}
\def\mn@doi@[#1]#2{\def\@tempa{#1}\ifx\@tempa\@empty \href
  {http://dx.doi.org/#2} {doi:#2}\else \href {http://dx.doi.org/#2} {#1}\fi
  \endgroup}
\def\mn@eprint#1#2{\mn@eprint@#1:#2::\@nil}
\def\mn@eprint@arXiv#1{\href {http://arxiv.org/abs/#1} {{\tt arXiv:#1}}}
\def\mn@eprint@dblp#1{\href {http://dblp.uni-trier.de/rec/bibtex/#1.xml}
  {dblp:#1}}
\def\mn@eprint@#1:#2:#3:#4\@nil{\def\@tempa {#1}\def\@tempb {#2}\def\@tempc
  {#3}\ifx \@tempc \@empty \let \@tempc \@tempb \let \@tempb \@tempa \fi \ifx
  \@tempb \@empty \def\@tempb {arXiv}\fi \@ifundefined
  {mn@eprint@\@tempb}{\@tempb:\@tempc}{\expandafter \expandafter \csname
  mn@eprint@\@tempb\endcsname \expandafter{\@tempc}}}

\bibitem[\protect\citeauthoryear{Adelberger}{Adelberger}{2001}]{Adelberger2001}
Adelberger E.~G.,  2001, \mn@doi [Classical and Quantum Gravity]
  {10.1088/0264-9381/18/13/302}, 18, 2397

\bibitem[\protect\citeauthoryear{{Adelberger}, {Fischbach}, {Krause}  \&
  {Newman}}{{Adelberger} et~al.}{2003}]{2003PhRvD..68f2002A}
{Adelberger} E.~G.,  {Fischbach} E.,  {Krause} D.~E.,   {Newman} R.~D.,  2003,
  \mn@doi [Physical Review D] {10.1103/PhysRevD.68.062002}, \href
  {http://cdsads.u-strasbg.fr/abs/2003PhRvD..68f2002A} {68, 062002}

\bibitem[\protect\citeauthoryear{{Altamimi}, {Rebischung}, {Métivier}  \&
  {Collilieux}}{{Altamimi} et~al.}{2016}]{2016JGRE..121.6109L}
{Altamimi} Z.,  {Rebischung} P.,  {Métivier} L.,   {Collilieux} X.,  2016,
  \mn@doi [Journal of Geophysical Research (Solid Earth)] {10.1002/jgre.20103},
  \href {http://cdsads.u-strasbg.fr/abs/2013JGRE..118.1435L} {121, 6109Ð6131}

\bibitem[\protect\citeauthoryear{{Anderson}, {Gross}, {Nordtvedt}  \&
  {Turyshev}}{{Anderson} et~al.}{1996}]{1996ApJ...459..365A}
{Anderson} J.~D.,  {Gross} M.,  {Nordtvedt} K.~L.,   {Turyshev} S.~G.,  1996,
  \mn@doi [\apj] {10.1086/176899}, \href
  {http://cdsads.u-strasbg.fr/abs/1996ApJ...459..365A} {459, 365}

\bibitem[\protect\citeauthoryear{Bender et~al.,}{Bender
  et~al.}{1973}]{Bender1973}
Bender P.~L.,  et~al., 1973, \mn@doi [Science] {10.1126/science.182.4109.229},
  182, 229

\bibitem[\protect\citeauthoryear{{Berti} et~al.,}{{Berti}
  et~al.}{2015}]{berti2015}
{Berti} E.,  et~al., 2015, \mn@doi [Classical and Quantum Gravity]
  {10.1088/0264-9381/32/24/243001}, \href
  {http://adsabs.harvard.edu/abs/2015CQGra..32x3001B} {32, 243001}

\bibitem[\protect\citeauthoryear{{Capozziello} \& {de Laurentis}}{{Capozziello}
  \& {de Laurentis}}{2011}]{capozziello2011}
{Capozziello} S.,  {de Laurentis} M.,  2011, \mn@doi [Physics Reports]
  {10.1016/j.physrep.2011.09.003}, \href
  {http://adsabs.harvard.edu/abs/2011PhR...509..167C} {509, 167}

\bibitem[\protect\citeauthoryear{Courde et~al.,}{Courde
  et~al.}{2017}]{Courde2017}
Courde C.,  et~al., 2017, \mn@doi [Astronomy {\&} Astrophysics]
  {10.1051/0004-6361/201628590}

\bibitem[\protect\citeauthoryear{{Damour} \& {Donoghue}}{{Damour} \&
  {Donoghue}}{2010}]{Damour2010}
{Damour} T.,  {Donoghue} J.~F.,  2010, \mn@doi [Physical Review D]
  {10.1103/PhysRevD.82.084033}, \href
  {http://adsabs.harvard.edu/abs/2010PhRvD..82h4033D} {82, 084033}

\bibitem[\protect\citeauthoryear{{Damour} \& {Polyakov}}{{Damour} \&
  {Polyakov}}{1994}]{damour:1994np}
{Damour} T.,  {Polyakov} A.~M.,  1994, \mn@doi [Nuclear Physics B]
  {10.1016/0550-3213(94)90143-0}, \href
  {http://adsabs.harvard.edu/abs/1994NuPhB.423..532D} {423, 532}

\bibitem[\protect\citeauthoryear{Donoghue}{Donoghue}{2017}]{donoghue2017}
Donoghue J.~F.,  2017, \mn@doi [Scholarpedia] {10.4249/scholarpedia.32997}, 12,
  32997

\bibitem[\protect\citeauthoryear{Faller, Winer, Carrion, Johnson, Spadin,
  Robinson, Wampler  \& Wieber}{Faller et~al.}{1969}]{Faller1969}
Faller J.,  Winer I.,  Carrion W.,  Johnson T.~S.,  Spadin P.,  Robinson L.,
  Wampler E.~J.,   Wieber D.,  1969, \mn@doi [Science]
  {10.1126/science.166.3901.99}, 166, 99

\bibitem[\protect\citeauthoryear{{Fienga}, {Laskar}, {Kuchynka}, {Manche},
  {Desvignes}, {Gastineau}, {Cognard}  \& {Theureau}}{{Fienga}
  et~al.}{2011}]{fienga:2011cm}
{Fienga} A.,  {Laskar} J.,  {Kuchynka} P.,  {Manche} H.,  {Desvignes} G.,
  {Gastineau} M.,  {Cognard} I.,   {Theureau} G.,  2011, \mn@doi [Celestial
  Mechanics and Dynamical Astronomy] {10.1007/s10569-011-9377-8}, \href
  {http://adsabs.harvard.edu/abs/2011CeMDA.111..363F} {111, 363}

\bibitem[\protect\citeauthoryear{Fienga, Manche, Laskar, Gastineau  \&
  Verma}{Fienga et~al.}{2014a}]{inpop13c}
Fienga A.,  Manche H.,  Laskar J.,  Gastineau M.,   Verma A.,  2014a, {INPOP
  new release: INPOP13c}

\bibitem[\protect\citeauthoryear{Fienga, Manche, Laskar, Gastineau  \&
  Verma}{Fienga et~al.}{2014b}]{Fienga2014}
Fienga A.,  Manche H.,  Laskar J.,  Gastineau M.,   Verma A.,  2014b, preprint,
  423 (\mn@eprint {arXiv} {1405.0484})

\bibitem[\protect\citeauthoryear{{Fienga}, {Laskar}, {Manche}  \&
  {Gastineau}}{{Fienga} et~al.}{2016}]{Fienga2016a}
{Fienga} A.,  {Laskar} J.,  {Manche} H.,   {Gastineau} M.,  2016, preprint,
  \href {http://adsabs.harvard.edu/abs/2016arXiv160100947F} {} (\mn@eprint
  {arXiv} {1601.00947})

\bibitem[\protect\citeauthoryear{{Folkner}, {Williams}  \& {Boggs}}{{Folkner}
  et~al.}{2009}]{Folkner2009}
{Folkner} W.~M.,  {Williams} J.~G.,   {Boggs} D.~H.,  2009, Interplanetary
  Network Progress Report, \href
  {http://adsabs.harvard.edu/abs/2009IPNPR.178C...1F} {178, 1}

\bibitem[\protect\citeauthoryear{{Folkner}, {Williams}, {Boggs}, {Park}  \&
  {Kuchynka}}{{Folkner} et~al.}{2014}]{Folkner2014}
{Folkner} W.~M.,  {Williams} J.~G.,  {Boggs} D.~H.,  {Park} R.~S.,   {Kuchynka}
  P.,  2014, Interplanetary Network Progress Report, \href
  {http://adsabs.harvard.edu/abs/2014IPNPR.196C...1F} {196, 1}

\bibitem[\protect\citeauthoryear{{Hees} \& {Minazzoli}}{{Hees} \&
  {Minazzoli}}{2015}]{hees:2015ax}
{Hees} A.,  {Minazzoli} O.,  2015, preprint, \href
  {http://adsabs.harvard.edu/abs/2015arXiv151205233H} {} (\mn@eprint {arXiv}
  {1512.05233})

\bibitem[\protect\citeauthoryear{Hofmann \& M{\"{u}}ller}{Hofmann \&
  M{\"{u}}ller}{2016}]{Hofmann2016}
Hofmann F.,  M{\"{u}}ller J.,  2016, in 20th International Workshop on Laser
  Ranging. \url
  {https://cddis.nasa.gov/lw20/docs/2016/presentations/30-Hofmann_presentation.pdf}

\bibitem[\protect\citeauthoryear{{Jeon}, {Cho}, {Kwak}, {Chung}, {Park}, {Lee}
  \& {Kuzmicz-Cieslak}}{{Jeon} et~al.}{2011}]{Jeon2011}
{Jeon} H.~S.,  {Cho} S.,  {Kwak} Y.~S.,  {Chung} J.~K.,  {Park} J.~U.,  {Lee}
  D.~K.,   {Kuzmicz-Cieslak} M.,  2011, \mn@doi [\apss]
  {10.1007/s10509-010-0528-2}, \href
  {http://cdsads.u-strasbg.fr/abs/2011Ap%26SS.332..341J} {332, 341}

\bibitem[\protect\citeauthoryear{{Joyce}, {Jain}, {Khoury}  \&
  {Trodden}}{{Joyce} et~al.}{2015}]{joyce2015}
{Joyce} A.,  {Jain} B.,  {Khoury} J.,   {Trodden} M.,  2015, \mn@doi [Physics
  Reports] {10.1016/j.physrep.2014.12.002}, \href
  {http://adsabs.harvard.edu/abs/2015PhR...568....1J} {568, 1}

\bibitem[\protect\citeauthoryear{Konopliv et~al.,}{Konopliv
  et~al.}{2013}]{Konopliv2013}
Konopliv A.~S.,  et~al., 2013, \mn@doi [Journal of Geophysical Research E:
  Planets] {10.1002/jgre.20097}, 118, 1415

\bibitem[\protect\citeauthoryear{{Konopliv} et~al.,}{{Konopliv}
  et~al.}{2014}]{2014GeoRL..41.1452K}
{Konopliv} A.~S.,  et~al., 2014, \mn@doi [\grl] {10.1002/2013GL059066}, \href
  {http://cdsads.u-strasbg.fr/abs/2014GeoRL..41.1452K} {41, 1452}

\bibitem[\protect\citeauthoryear{{Kosteleck{\'y}}}{{Kosteleck{\'y}}}{2004}]{kostelecky2004}
{Kosteleck{\'y}} V.~A.,  2004, \mn@doi [Physical Review D]
  {10.1103/PhysRevD.69.105009}, \href
  {http://adsabs.harvard.edu/abs/2004PhRvD..69j5009K} {69, 105009}

\bibitem[\protect\citeauthoryear{{Liorzou}, {Boulanger}, {Rodrigues}, {Touboul}
   \& {Selig}}{{Liorzou} et~al.}{2014}]{2014AdSpR..54.1119L}
{Liorzou} F.,  {Boulanger} D.,  {Rodrigues} M.,  {Touboul} P.,   {Selig} H.,
  2014, \mn@doi [Advances in Space Research] {10.1016/j.asr.2014.05.009}, \href
  {http://cdsads.u-strasbg.fr/abs/2014AdSpR..54.1119L} {54, 1119}

\bibitem[\protect\citeauthoryear{{Lucchesi}, {Anselmo}, {Bassan}, {Pardini},
  {Peron}, {Pucacco}  \& {Visco}}{{Lucchesi}
  et~al.}{2015}]{2015CQGra..32o5012L}
{Lucchesi} D.~M.,  {Anselmo} L.,  {Bassan} M.,  {Pardini} C.,  {Peron} R.,
  {Pucacco} G.,   {Visco} M.,  2015, \mn@doi [Classical and Quantum Gravity]
  {10.1088/0264-9381/32/15/155012}, \href
  {http://cdsads.u-strasbg.fr/abs/2015CQGra..32o5012L} {32, 155012}

\bibitem[\protect\citeauthoryear{Lyard, Lefevre, Letellier  \& Francis}{Lyard
  et~al.}{2006}]{Lyard2006}
Lyard F.,  Lefevre F.,  Letellier T.,   Francis O.,  2006, \mn@doi [Ocean
  Dynamics] {10.1007/s10236-006-0086-x}, 56, 394

\bibitem[\protect\citeauthoryear{Manche}{Manche}{2011}]{Manche2011}
Manche H.,  2011, Phd dissertation, Observatoire de Paris, \url
  {https://tel.archives-ouvertes.fr/tel-00689852}

\bibitem[\protect\citeauthoryear{Marty et~al.,}{Marty
  et~al.}{2011}]{marty2011gins}
Marty J.,  et~al., 2011, in 3rd International Colloquium Scientific and
  Fundamental Aspects of the Galileo Programme, ESA Proceedings WPP326. \url
  {http://hpiers.obspm.fr/combinaison/documentation/articles/GINS_Marty.pdf}

\bibitem[\protect\citeauthoryear{{Matsumoto}, {Yamada}, {Kikuchi}, {Kamata},
  {Ishihara}, {Iwata}, {Hanada}  \& {Sasaki}}{{Matsumoto}
  et~al.}{2015}]{2015GeoRL..42.7351M}
{Matsumoto} K.,  {Yamada} R.,  {Kikuchi} F.,  {Kamata} S.,  {Ishihara} Y.,
  {Iwata} T.,  {Hanada} H.,   {Sasaki} S.,  2015, \mn@doi [\grl]
  {10.1002/2015GL065335}, \href
  {http://cdsads.u-strasbg.fr/abs/2015GeoRL..42.7351M} {42, 7351}

\bibitem[\protect\citeauthoryear{{Matsuo}, {Chao}, {Otsubo}  \&
  {Heki}}{{Matsuo} et~al.}{2013}]{2013GeoRL..40.4662M}
{Matsuo} K.,  {Chao} B.~F.,  {Otsubo} T.,   {Heki} K.,  2013, \mn@doi [\grl]
  {10.1002/grl.50900}, \href
  {http://cdsads.u-strasbg.fr/abs/2013GeoRL..40.4662M} {40, 4662}

\bibitem[\protect\citeauthoryear{{Matsuyama}, {Nimmo}, {Keane}, {Chan},
  {Taylor}, {Wieczorek}, {Kiefer}  \& {Williams}}{{Matsuyama}
  et~al.}{2016}]{2016GeoRL..43.8365M}
{Matsuyama} I.,  {Nimmo} F.,  {Keane} J.~T.,  {Chan} N.~H.,  {Taylor} G.~J.,
  {Wieczorek} M.~A.,  {Kiefer} W.~S.,   {Williams} J.~G.,  2016, \mn@doi [\grl]
  {10.1002/2016GL069952}, \href
  {http://cdsads.u-strasbg.fr/abs/2016GeoRL..43.8365M} {43, 8365}

\bibitem[\protect\citeauthoryear{Mazarico, Barker, Neumann, Zuber  \&
  Smith}{Mazarico et~al.}{2014}]{Mazarico2014}
Mazarico E.,  Barker M.~K.,  Neumann G.~A.,  Zuber M.~T.,   Smith D.~E.,  2014,
  \mn@doi [Geophysical Research Letters] {10.1002/2013GL059085}, 41, 2282

\bibitem[\protect\citeauthoryear{{McCarthy} \& {Petit}}{{McCarthy} \&
  {Petit}}{2004}]{McCarthy2004}
{McCarthy} D.~D.,  {Petit} G.,  2004, IERS Technical Note, \href
  {http://adsabs.harvard.edu/abs/2004ITN....32....1M} {32}

\bibitem[\protect\citeauthoryear{{Merkowitz}}{{Merkowitz}}{2010}]{2010LRR....13....7M}
{Merkowitz} S.~M.,  2010, Living Reviews in Relativity, \href
  {http://cdsads.u-strasbg.fr/abs/2010LRR....13....7M} {13, 7}

\bibitem[\protect\citeauthoryear{{Minazzoli} \& {Hees}}{{Minazzoli} \&
  {Hees}}{2016}]{minazzoli:2016pr}
{Minazzoli} O.,  {Hees} A.,  2016, \mn@doi [Physical Review D]
  {10.1103/PhysRevD.94.064038}, \href
  {http://adsabs.harvard.edu/abs/2016PhRvD..94f4038M} {94, 064038}

\bibitem[\protect\citeauthoryear{{Minazzoli}, {Bernus}, {Fienga}, {Hees},
  {Laskar}  \& {Viswanathan}}{{Minazzoli} et~al.}{2017}]{2017arXiv170505244M}
{Minazzoli} O.,  {Bernus} L.,  {Fienga} A.,  {Hees} A.,  {Laskar} J.,
  {Viswanathan} V.,  2017, preprint, \href
  {http://adsabs.harvard.edu/abs/2017arXiv170505244M} {} (\mn@eprint {arXiv}
  {1705.05244})

\bibitem[\protect\citeauthoryear{Moyer}{Moyer}{2003}]{Moyer2003}
Moyer T.~D.,  2003, {Formulation for Observed and Computed Values of Deep Space
  Network Data Types for Navigation}.
~ Vol. 2, John Wiley {\&} Sons, Inc., Hoboken, NJ, USA,
  \mn@doi{10.1002/0471728470}

\bibitem[\protect\citeauthoryear{M{\"{u}}ller, Hofmann  \&
  Biskupek}{M{\"{u}}ller et~al.}{2012}]{Muller2012}
M{\"{u}}ller J.,  Hofmann F.,   Biskupek L.,  2012, \mn@doi [Classical and
  Quantum Gravity] {10.1088/0264-9381/29/18/184006}, 29, 184006

\bibitem[\protect\citeauthoryear{Murphy}{Murphy}{2013}]{Murphy2013a}
Murphy T.~W.,  2013, \mn@doi [Reports on Progress in Physics]
  {10.1088/0034-4885/76/7/076901}, 76, 076901

\bibitem[\protect\citeauthoryear{Murphy et~al.,}{Murphy
  et~al.}{2011}]{Murphy2011}
Murphy T.~W.,  et~al., 2011, \mn@doi [Icarus] {10.1016/j.icarus.2010.11.010},
  211, 1103

\bibitem[\protect\citeauthoryear{Murphy, Adelberger, Battat, Hoyle, Johnson,
  McMillan, Stubbs  \& Swanson}{Murphy et~al.}{2012}]{Murphy2012}
Murphy T.~W.,  Adelberger E.~G.,  Battat J. B.~R.,  Hoyle C.~D.,  Johnson
  N.~H.,  McMillan R.~J.,  Stubbs C.~W.,   Swanson H.~E.,  2012, \mn@doi
  [Classical and Quantum Gravity] {10.1088/0264-9381/29/18/184005}, 29, 184005

\bibitem[\protect\citeauthoryear{Murphy, McMillan, Johnson  \& Goodrow}{Murphy
  et~al.}{2014}]{Murphy2014}
Murphy T.~W.,  McMillan R.~J.,  Johnson N.~H.,   Goodrow S.~D.,  2014, \mn@doi
  [Icarus] {10.1016/j.icarus.2013.12.006}, 231, 183

\bibitem[\protect\citeauthoryear{{Nordtvedt}}{{Nordtvedt}}{1968a}]{1968PhRv..169.1017N}
{Nordtvedt} K.,  1968a, \mn@doi [Physical Review] {10.1103/PhysRev.169.1017},
  \href {http://cdsads.u-strasbg.fr/abs/1968PhRv..169.1017N} {169, 1017}

\bibitem[\protect\citeauthoryear{{Nordtvedt}}{{Nordtvedt}}{1968b}]{1968PhRv..170.1186N}
{Nordtvedt} K.,  1968b, \mn@doi [Physical Review] {10.1103/PhysRev.170.1186},
  \href {http://cdsads.u-strasbg.fr/abs/1968PhRv..170.1186N} {170, 1186}

\bibitem[\protect\citeauthoryear{{Nordtvedt}}{{Nordtvedt}}{1998}]{1998CQGra..15.3363N}
{Nordtvedt} K.,  1998, \mn@doi [Classical and Quantum Gravity]
  {10.1088/0264-9381/15/11/005}, \href
  {http://cdsads.u-strasbg.fr/abs/1998CQGra..15.3363N} {15, 3363}

\bibitem[\protect\citeauthoryear{{Nordtvedt}}{{Nordtvedt}}{2014}]{nordtvedt:2014sc}
{Nordtvedt} K.,  2014, \mn@doi [Scholarpedia] {10.4249/scholarpedia.32141HTML:
  http://www.scholarpedia.org/article/Nordtvedt_effect}, \href
  {http://adsabs.harvard.edu/abs/2014SchpJ...932141N} {9, 32141}

\bibitem[\protect\citeauthoryear{Oberst et~al.,}{Oberst
  et~al.}{2012}]{Oberst2012}
Oberst J.,  et~al., 2012, \mn@doi [Planetary and Space Science]
  {10.1016/j.pss.2012.10.005}, 74, 179

\bibitem[\protect\citeauthoryear{Pavlis, Holmes, Kenyon  \& Factor}{Pavlis
  et~al.}{2012}]{Pavlis2012}
Pavlis N.~K.,  Holmes S.~A.,  Kenyon S.~C.,   Factor J.~K.,  2012, \mn@doi
  [Journal of Geophysical Research: Solid Earth] {10.1029/2011JB008916}, 117,
  n/a

\bibitem[\protect\citeauthoryear{Pavlis, Holmes, Kenyon  \& Factor}{Pavlis
  et~al.}{2013}]{Pavlis2013}
Pavlis N.~K.,  Holmes S.~A.,  Kenyon S.~C.,   Factor J.~K.,  2013, \mn@doi
  [Journal of Geophysical Research: Solid Earth] {10.1002/jgrb.50167,2013},
  118, 2633

\bibitem[\protect\citeauthoryear{Pavlov, Williams  \& Suvorkin}{Pavlov
  et~al.}{2016}]{Pavlov2016}
Pavlov D.~A.,  Williams J.~G.,   Suvorkin V.~V.,  2016, \mn@doi [Celestial
  Mechanics and Dynamical Astronomy] {10.1007/s10569-016-9712-1}, 126, 61

\bibitem[\protect\citeauthoryear{{Peron}}{{Peron}}{2013}]{2013MNRAS.432.2591P}
{Peron} R.,  2013, \mn@doi [\mnras] {10.1093/mnras/stt621}, \href
  {http://cdsads.u-strasbg.fr/abs/2013MNRAS.432.2591P} {432, 2591}

\bibitem[\protect\citeauthoryear{{Petit} \& {Luzum}}{{Petit} \&
  {Luzum}}{2010}]{IERS2010}
{Petit} G.,  {Luzum} B.,  2010, IERS Technical Note, \href
  {http://adsabs.harvard.edu/abs/2010ITN....36....1P} {36}

\bibitem[\protect\citeauthoryear{{Ries} et~al.,}{{Ries}
  et~al.}{2016}]{Ries2016}
{Ries} J.,  et~al., 2016, \mn@doi [{GFZ Data Services}]
  {10.5880/icgem.2016.002}

\bibitem[\protect\citeauthoryear{Samain et~al.,}{Samain
  et~al.}{1998}]{Samain1998}
Samain E.,  et~al., 1998, \mn@doi [Astronomy and Astrophysics Supplement
  Series] {10.1051/aas:1998227}, 130, 235

\bibitem[\protect\citeauthoryear{Standish \& Williams}{Standish \&
  Williams}{1992}]{standish1992orbital}
Standish E.~M.,  Williams J.~G.,  1992, Explanatory supplement to the
  astronomical almanac, \href
  {ftp://ssd.jpl.nasa.gov/pub/eph/planets/ioms/ExplSupplChap8.pdf} {pp
  279--323}

\bibitem[\protect\citeauthoryear{{Viswanathan}}{{Viswanathan}}{2017}]{Viswanathan2017b}
{Viswanathan} V.,  2017, Phd dissertation (submitted), Observatoire de Paris

\bibitem[\protect\citeauthoryear{{Viswanathan}, {Fienga}, {Laskar}, {Manche},
  {Torre}, {Courde}  \& {Exertier}}{{Viswanathan}
  et~al.}{2015}]{viswanathan2015utilizing}
{Viswanathan} V.,  {Fienga} A.,  {Laskar} J.,  {Manche} H.,  {Torre} J.-M.,
  {Courde} C.,   {Exertier} P.,  2015, IAU General Assembly, \href
  {http://adsabs.harvard.edu/abs/2015IAUGA..2228567V} {22, 2228567}

\bibitem[\protect\citeauthoryear{{Viswanathan} et~al.,}{{Viswanathan}
  et~al.}{2016}]{2016EGUGA..1813995V}
{Viswanathan} V.,  et~al., 2016, EGU General Assembly Conference Abstracts,
  \href {http://adsabs.harvard.edu/abs/2016EGUGA..1813995V} {18, 13995}

\bibitem[\protect\citeauthoryear{{Viswanathan}, {Fienga}, {Gastineau}  \&
  {Laskar}}{{Viswanathan} et~al.}{2017}]{Viswanathan2017a}
{Viswanathan} V.,  {Fienga} A.,  {Gastineau} M.,   {Laskar} J.,  2017, Notes
  Scientifiques et Techniques de l'Institut de M{\'e}canique C{\'e}leste, \href
  {http://adsabs.harvard.edu/abs/2017NSTBL.108.....V} {108, 1}

\bibitem[\protect\citeauthoryear{Vokrouhlick{\'{y}}}{Vokrouhlick{\'{y}}}{1997}]{Vokrouhlicky1997}
Vokrouhlick{\'{y}} D.,  1997, \mn@doi [Icarus] {10.1006/icar.1996.5652}, 126,
  293

\bibitem[\protect\citeauthoryear{Wieczorek}{Wieczorek}{2007}]{Wieczorek2007}
Wieczorek M.,  2007, in Schubert G.,  ed., , Treatise on Geophysics.
Elsevier, Amsterdam, pp 165 -- 206, \mn@doi{10.1016/B978-044452748-6.00156-5}

\bibitem[\protect\citeauthoryear{{Will}}{{Will}}{2014}]{will:2014lr}
{Will} C.~M.,  2014, \mn@doi [Living Reviews in Relativity]
  {10.12942/lrr-2014-4}, \href
  {http://adsabs.harvard.edu/abs/2014LRR....17....4W} {17, 4}

\bibitem[\protect\citeauthoryear{Williams \& Boggs}{Williams \&
  Boggs}{2015}]{Williams2015}
Williams J.~G.,  Boggs D.~H.,  2015, \mn@doi [Journal of Geophysical Research:
  Planets] {10.1002/2014JE004755}, 120, 689

\bibitem[\protect\citeauthoryear{Williams \& Boggs}{Williams \&
  Boggs}{2016}]{Williams2016Secular}
Williams J.~G.,  Boggs D.~H.,  2016, \mn@doi [Celestial Mechanics and Dynamical
  Astronomy] {10.1007/s10569-016-9702-3}, 126, 89

\bibitem[\protect\citeauthoryear{Williams, Boggs, Yoder, Ratcliff  \&
  Dickey}{Williams et~al.}{2001}]{Williams2001}
Williams J.~G.,  Boggs D.~H.,  Yoder C.~F.,  Ratcliff J.~T.,   Dickey J.~O.,
  2001, \mn@doi [Journal of Geophysical Research] {10.1029/2000JE001396}, 106,
  27933

\bibitem[\protect\citeauthoryear{Williams, Turyshev  \& Boggs}{Williams
  et~al.}{2009}]{Williams2009}
Williams J.,  Turyshev S.,   Boggs D.,  2009, \mn@doi [International Journal of
  Modern Physics D] {10.1142/S021827180901500X}, 18, 1129

\bibitem[\protect\citeauthoryear{Williams, Turyshev  \& Boggs}{Williams
  et~al.}{2012}]{Williams2012}
Williams J.~G.,  Turyshev S.~G.,   Boggs D.~H.,  2012, \mn@doi [Classical and
  Quantum Gravity] {10.1088/0264-9381/29/18/184004}, 29, 184004

\bibitem[\protect\citeauthoryear{{Woodard}}{{Woodard}}{2009}]{woodard2009}
{Woodard} R.~P.,  2009, \mn@doi [Reports on Progress in Physics]
  {10.1088/0034-4885/72/12/126002}, \href
  {http://adsabs.harvard.edu/abs/2009RPPh...72l6002W} {72, 126002}

\bibitem[\protect\citeauthoryear{Yunes, Yagi  \& Pretorius}{Yunes
  et~al.}{2016}]{Yunes2016}
Yunes N.,  Yagi K.,   Pretorius F.,  2016, \mn@doi [Physical Review D]
  {10.1103/PhysRevD.94.084002}, 94, 084002

\makeatother
\end{thebibliography}

\label{lastpage}
\end{document}